\documentclass[prd,aps,showpacs,epsf,floats,onecolumn]{revtex4}%
\usepackage{amssymb}
\usepackage{amsfonts}
\usepackage{amsmath}
\usepackage{graphicx}%
\providecommand{\U}[1]{\protect\rule{.1in}{.1in}}
\begin{document}
\title{\textbf{Information Geometric Perspective on Off-Resonance Effects in Driven
Two-Level Quantum Systems}}
\author{\textbf{Carlo Cafaro}$^{1}$, \textbf{Steven Gassner}$^{2}$, and \textbf{Paul
M.\ Alsing}$^{3}$}
\affiliation{$^{1,2}$SUNY Polytechnic Institute, 12203 Albany, New York, USA}
\affiliation{$^{3}$Air Force Research Laboratory, Information Directorate, 13441 Rome, New
York, USA}

\begin{abstract}
We present an information geometric analysis of off-resonance effects on
classes of exactly solvable generalized semi-classical Rabi systems.
Specifically, we consider population transfer performed by four distinct
off-resonant driving schemes specified by $\mathrm{su}\left(  2\text{; }%
\mathbb{C}
\right)  $ time-dependent Hamiltonian models. For each scheme, we study the
consequences of a departure from the on-resonance condition in terms of both
geodesic paths and geodesic speeds on the corresponding manifold of transition
probability vectors. In particular, we analyze the robustness of each driving
scheme against off-resonance effects. Moreover, we report on a possible
tradeoff between speed and robustness in the driving schemes being
investigated. Finally, we discuss the emergence of a different relative
ranking in terms of performance among the various driving schemes when
transitioning from on-resonant to off-resonant scenarios.

\end{abstract}

\pacs{Information Theory (89.70.+c), Probability Theory (02.50.Cw), Quantum
Mechanics (03.65.-w), Riemannian Geometry (02.40.Ky), Statistical Mechanics (05.20.-y).}
\maketitle

\section{Introduction}

It is well-known that the resonance phenomenon concerns the amplification that
occurs when the frequency of an externally applied field (for instance, a
magnetic field) is equal to a natural characteristic frequency of the system
on which it acts. When a field is applied at the resonant frequency of another
system, the system oscillates at a higher amplitude than when the field is
applied at a non-resonant frequency. In an actual experimental laboratory,
off-resonance effects can occur for a number of reasons. For instance, a
common source of off-resonance effects is the presence of unforeseen magnetic
field gradients that are not part of the externally applied magnetic field.
These gradients can cause signal loss and, as a consequence, signals at the
\textquotedblleft wrong\textquotedblright\ resonant frequency (that is,
off-resonance). Furthermore, when the intensity of an applied magnetic field
gradient is changed in a laboratory, decaying eddy currents can produce
time-varying off-resonant effects \cite{jackson}. Additionally, the presence
of concomitant gradients can create off-resonance effects in the form of
$xy$-spatial variations in the Larmor frequency when a magnetic field gradient
in the $z$-direction is applied \cite{sakurai}. For the interested reader, we
discuss two illustrative examples concerning the departure from the
on-resonance condition in the context of classical and quantum mechanics in
Appendix A.

The resonance phenomenon happens to play a key role in problems where one
seeks to drive an initial source state towards a final target state. For
instance, typical problems of this kind can be found when controlling
population transfer in quantum systems \cite{dalessandro01,boscain02,romano14}
or when searching for a target state \cite{byrnes18,carloIJQI}. In the
framework of quantum control theory, it is commonly believed that an
off-resonant driving scheme cannot help achieving population transfer with
high fidelity between two quantum states. However, there do exist theoretical
investigations relying on the phenomenon of superoscillations
\cite{aharonov90} where population transfer occurs by using only off-resonant
driving fields \cite{kempf17}. In particular, during resonant
superoscillations, the quantum system is temporarily excited and behaves as if
truly driven at resonance \cite{kempf17}. Furthermore, the existence of
suboptimal transfer efficiencies, caused by the sequential use of two lasers
in order to physically realize a transition from highly excited Rydberg states
to ground states \cite{barredo15,schempp15}, leads to question the possibility
of transition from a ground state to the Rydberg state by means of lasers
far-off-resonant with respect to the transition frequency. Indeed, motivated
by such possibility, driving schemes specified by periodic square-wells and
Gaussian pulses in order to implement coherent transfer with far-off-resonant
driving fields have been proposed in Ref. \cite{shi16}.

It is also well-known that on-resonance driving is not a good control strategy
for a quantum system if, from an experimental standpoint, one allows the Rabi
frequency (that is, the frequency that describes the coupling between the
driving field and the two-level system) to be higher than the Larmor frequency
(that is, the characteristic transition frequency of the two-level system in
the absence of the external driving field) \cite{cappellaro18}. This is called
the strong driving regime. In such a scenario, the so-called rotating wave
approximation (RWA, \cite{eberly87}) is not applicable and the system is
driven with an external field whose amplitude is greater than or equal to the
energy splitting between the system's states. For an analysis on the deviation
from sinusoidal Rabi oscillations by studying the time-evolution dynamics of a
strongly driven dressed electron spin in silicon, we refer to Ref.
\cite{laucht16}. Furthermore, for an explicit manifestation of strong
sensitivity to the initial phase of the driving field in the dynamics of a
general semi-classical Rabi model in regimes of arbitrary strong driving, we
refer to Ref. \cite{dai17}.

Based upon our considerations, it is clear that off-resonance phenomena
exhibit both experimental (for instance, limited control of spurious
electromagnetic signals and strong driving regimes) and theoretical (for
instance, sensitivity to initial conditions and emergence of more complex
dynamical scenarios) interest. In the framework of analog quantum searching,
we have recently presented a detailed investigation concerning the physical
connection between quantum search Hamiltonians and exactly solvable
time-dependent two-level quantum systems in Ref. \cite{carloIJQI}. In Refs.
\cite{carlophysica, gassner20}, instead, we analyzed the possibility of
modifying the original Farhi-Gutmann Hamiltonian quantum search algorithm
\cite{farhi98} in order to speed up the procedure for producing a suitably
distributed unknown normalized quantum mechanical state provided only a nearly
optimal fidelity is sought. In Ref. \cite{carlopre}, we presented an
information geometric characterization of the oscillatory or monotonic
behavior of statistically parametrized squared probability amplitudes
originating from special functional forms of the Fisher information function:
constant, exponential decay, and power-law decay. Furthermore, for each case,
we computed both the computational speed and the availability loss of the
corresponding physical processes by exploiting a convenient Riemannian
geometrization of useful thermodynamical concepts. Finally, building upon our
works presented in Refs. \cite{carloIJQI,carlophysica,gassner20,carlopre}, we
presented in Ref. \cite{carlotechnical19} an information geometric analysis of
geodesic speeds and entropy production rates in geodesic motion on manifolds
of parametrized quantum states. These pure states emerged as outputs of
suitable $\mathrm{su}\left(  2\text{; }%
\mathbb{C}
\right)  $ time-dependent Hamiltonian evolutions used to describe distinct
types of analog quantum search schemes viewed as driving strategies. In
particular, by evaluating the\textbf{\ }geodesic speed and the total entropy
production along the optimum transfer paths in a number of physical scenarios
of interest in analog quantum search problems (for instance, constant,
oscillatory, power-law decay, and exponential decay of the driving magnetic
fields), we showed in an explicit quantitative manner that to a faster
transfer there corresponds necessarily a higher entropy production rate. Thus,
we concluded that lower entropic efficiency values do appear to accompany
higher entropic geodesic speed values in quantum transfer processes. Our
information geometric analysis in Ref. \cite{carlotechnical19} was limited to
the case in which the on-resonance driving condition was satisfied for all
quantum driving strategies. However, based upon our previous considerations,
it would be of theoretical interest to investigate from an information
geometric perspective the effects of off-resonance effects on these driving
strategies. In particular, we would like to address the following questions:

\begin{itemize}
\item[(i)] What are the main effects of deviations from the on-resonance
condition in these $\mathrm{su}\left(  2\text{; }%
\mathbb{C}
\right)  $ time-dependent Hamiltonian evolutions?

\item[(ii)] Do off-resonance effects modify (with respect to the on-resonance
scenario) the relative ranking in terms of performance (quantified in terms of
geodesic speed and/or minimum transfer time) among the driving schemes being considered?

\item[(iii)] Are there driving schemes that are especially robust against
deviations from the on-resonance condition and that, in addition, are capable
of reaching\textbf{ }sufficiently high fidelity values?
\end{itemize}

The rest of the paper is organized as follows. In Section II, we present our
driving strategies in terms of classes of exactly solvable generalized
semi-classical Rabi systems. Specifically, we consider the population transfer
performed by four distinct off-resonant driving schemes specified by
$\mathrm{su}\left(  2\text{; }%
\mathbb{C}
\right)  $ time-dependent Hamiltonian models. In Section III, we propose our
information geometric analysis of off-resonance effects on such classes of
systems. For each scheme, we study the consequences of a departure from the
on-resonance condition in terms of both geodesic paths and geodesic speeds on
the corresponding manifold of transition probability vectors. In particular,
we analyze the robustness of each driving scheme against off-resonance
effects. Moreover, we report on a possible tradeoff between speed and
robustness in the driving schemes being investigated. Finally, we discuss the
emergence of a different relative ranking in terms of performance among the
various driving schemes when transitioning from the on-resonant to the
off-resonant scenarios. Our conclusive remarks appear in Section IV. Auxiliary
illustrative examples together with some technical details are placed in
Appendices A and B.

\section{The \textrm{su}$(2$; $%
\mathbb{C}
)$ quantum driving}

In this section, we introduce our driving strategies in terms of classes of
exactly solvable generalized semi-classical Rabi systems
\cite{messina14,grimaudo18}.

The quantum evolution we take into consideration is specified by means of an
Hamiltonian operator $\mathcal{H}_{\mathrm{su}\left(  2\text{; }%
\mathbb{C}
\right)  }$ written as the most general linear superposition of the three
traceless and anti-Hermitian generators $\left\{  i\sigma_{x}\text{, }%
-i\sigma_{y}\text{, }i\sigma_{z}\right\}  $ of $\mathrm{su}\left(  2\text{; }%
\mathbb{C}
\right)  $, the Lie algebra of the special unitary group $\mathrm{SU}\left(
2\text{; }%
\mathbb{C}
\right)  $,%
\begin{equation}
\mathcal{H}_{\mathrm{su}\left(  2\text{; }%
\mathbb{C}
\right)  }\left(  t\right)  \overset{\text{def}}{=}a\left(  t\right)  \left(
i\sigma_{x}\right)  +b\left(  t\right)  \left(  -i\sigma_{y}\right)  +c\left(
t\right)  \left(  \text{ }i\sigma_{z}\right)  \text{.} \label{easy}%
\end{equation}
In Eq. (\ref{easy}), $a\left(  t\right)  $, $b\left(  t\right)  $, and
$c\left(  t\right)  $ are time-dependent complex coefficients while
$\vec{\sigma}\overset{\text{def}}{=}\left(  \sigma_{x}\text{, }\sigma
_{y}\text{, }\sigma_{z}\right)  $ denotes the Pauli vector operator. In
particular, by putting $a\left(  t\right)  \overset{\text{def}}{=}-i\omega
_{x}\left(  t\right)  $, $b\left(  t\right)  \overset{\text{def}}{=}%
i\omega_{y}\left(  t\right)  $, and $c\left(  t\right)  \overset{\text{def}%
}{=}-i\Omega\left(  t\right)  $, Eq. (\ref{easy}) yields%
\begin{equation}
\mathcal{H}_{\mathrm{su}\left(  2\text{; }%
\mathbb{C}
\right)  }\left(  t\right)  \overset{\text{def}}{=}\omega_{x}\left(  t\right)
\sigma_{x}+\omega_{y}\left(  t\right)  \sigma_{y}+\Omega\left(  t\right)
\sigma_{z}\text{.} \label{easy2}%
\end{equation}
In the framework of $\mathrm{su}\left(  2\text{; }%
\mathbb{C}
\right)  $ Hamiltonian models, $\omega\left(  t\right)  \overset{\text{def}%
}{=}\omega_{x}\left(  t\right)  -i\omega_{y}\left(  t\right)  =\omega
_{\mathcal{H}}\left(  t\right)  e^{i\phi_{\omega}\left(  t\right)  }$ and
$\Omega\left(  t\right)  $ are the so-called complex transverse field and real
longitudinal field, respectively. Obviously, $\omega_{\mathcal{H}}\left(
t\right)  $ denotes the modulus of $\omega\left(  t\right)  $. In what
follows, we assume that longitudinal fields $\Omega\left(  t\right)  $ are
oriented\textbf{\ }along the $z$-axis while transverse fields $\omega\left(
t\right)  $ lie in the $xy$-plane. Taking into consideration the quantum
evolution of an electron (or, more generally, a spin\textbf{-}$1/2$\textbf{
}particle\textbf{)} in an external time-dependent magnetic field $\vec
{B}\left(  t\right)  $, the Hamiltonian $\mathcal{H}_{\mathrm{su}\left(
2\text{; }%
\mathbb{C}
\right)  }\left(  t\right)  $ in Eq. (\ref{easy2}) can be rewritten as%
\begin{equation}
\mathcal{H}_{\mathrm{su}\left(  2\text{; }%
\mathbb{C}
\right)  }\left(  t\right)  \overset{\text{def}}{=}-\vec{\mu}\cdot\vec
{B}\left(  t\right)  \text{,} \label{easy3}%
\end{equation}
where $\vec{\mu}\overset{\text{def}}{=}\left(  e\hslash/2mc\right)
\vec{\sigma}$ denotes the magnetic moment of the electron with $\mu
_{\text{Bohr}}\overset{\text{def}}{=}\left\vert e\right\vert \hslash
/(2mc)\simeq9.27\times10^{-21}\left[  \mathrm{cgs}\right]  $ being the
so-called Bohr magneton. The quantity $m\simeq9.11\times10^{-31}\left[
\text{\textrm{MKSA}}\right]  $\textbf{ }is the mass of an electron while
$\left\vert e\right\vert \simeq1.60\times10^{-19}\left[  \text{\textrm{MKSA}%
}\right]  $ denotes\textbf{ }the absolute value of the electric charge of an
electron. Moreover, $c\simeq3\times10^{8}\left[  \text{\textrm{MKSA}}\right]
$ and $\hslash\simeq1.05\times10^{-34}\left[  \text{\textrm{MKSA}}\right]  $
are\textbf{ }the speed of light and the reduced Planck constant, respectively.
The magnetic field $\vec{B}\left(  t\right)  $ in Eq. (\ref{easy}) can be
recast as
\begin{equation}
\vec{B}\left(  t\right)  \overset{\text{def}}{=}\vec{B}_{\perp}\left(
t\right)  +\vec{B}_{\parallel}\left(  t\right)  \text{,} \label{easy4}%
\end{equation}
with $\vec{B}_{\perp}\left(  t\right)  \overset{\text{def}}{=}B_{x}\left(
t\right)  \hat{x}+B_{y}\left(  t\right)  \hat{y}$ and $\vec{B}_{\parallel
}\left(  t\right)  \overset{\text{def}}{=}B_{z}\left(  t\right)  \hat{z}$.
Equating\textbf{ }the Hamiltonians in Eqs. (\ref{easy2}) and (\ref{easy3}) and
making use of\textbf{ }the magnetic field decomposition in Eq. (\ref{easy4}),
the connection between the set of field intensities $\left\{  \omega
_{\mathcal{H}}\left(  t\right)  \text{, }\Omega_{\mathcal{H}}\left(  t\right)
\right\}  $ and the set of magnetic field intensities $\left\{  B_{\perp
}\left(  t\right)  \text{, }B_{\parallel}\left(  t\right)  \right\}  $ becomes
clear. More specifically, we observe that $B_{\perp}\left(  t\right)
\propto\omega_{\mathcal{H}}\left(  t\right)  $ and $B_{\parallel}\left(
t\right)  \propto$ $\Omega_{\mathcal{H}}\left(  t\right)  \overset{\text{def}%
}{=}\left\vert \Omega\left(  t\right)  \right\vert $. The exact connection
between the components $\left\{  B_{x}\left(  t\right)  \text{, }B_{y}\left(
t\right)  \text{, }B_{z}\left(  t\right)  \right\}  $ and $\left\{  \omega
_{x}\left(  t\right)  \text{, }\omega_{y}\left(  t\right)  \text{, }%
\Omega\left(  t\right)  \right\}  $ is expressed\textbf{ }by%
\begin{equation}
B_{x}\left(  t\right)  =-\frac{2mc}{e\hslash}\omega_{x}\left(  t\right)
\text{, }B_{y}\left(  t\right)  =-\frac{2mc}{e\hslash}\omega_{y}\left(
t\right)  \text{, and }B_{z}\left(  t\right)  =-\frac{2mc}{e\hslash}%
\Omega\left(  t\right)  \text{.}%
\end{equation}
Moreover, in terms of field intensities, we get%
\begin{equation}
B_{\perp}\left(  t\right)  =\frac{2mc}{\left\vert e\right\vert \hslash}%
\omega_{\mathcal{H}}\left(  t\right)  \text{, and }B_{\parallel}\left(
t\right)  =\frac{2mc}{\left\vert e\right\vert \hslash}\Omega_{\mathcal{H}%
}\left(  t\right)  \text{.}%
\end{equation}
Studying the quantum evolution of an electron governed by the Hamiltonian in
Eq. (\ref{easy3}) by means of exact analytical expressions of complex
probability amplitudes and/or real transition probabilities from an initial
source state to a final target state can be quite challenging. In particular,
the canonical matrix representations of $\mathcal{H}_{\mathrm{su}\left(
2\text{; }%
\mathbb{C}
\right)  }$ in Eq. (\ref{easy2}) and $\mathcal{U}\left(  t\right)  $, the
unitary evolution operator arising from $\mathcal{H}_{\mathrm{su}\left(
2\text{; }%
\mathbb{C}
\right)  }$, with $i\hslash\mathcal{\dot{U}}\left(  t\right)  =\mathcal{H}%
_{\mathrm{su}\left(  2\text{; }%
\mathbb{C}
\right)  }\mathcal{U}\left(  t\right)  $ and $\mathcal{\dot{U}}\overset
{\text{def}}{=}\partial_{t}\mathcal{U}$, are given by%
\begin{equation}
\left[  \mathcal{H}_{\mathrm{su}\left(  2\text{; }%
\mathbb{C}
\right)  }\right]  \overset{\text{def}}{=}\left(
\begin{array}
[c]{cc}%
\Omega\left(  t\right)  & \omega\left(  t\right) \\
\omega^{\ast}\left(  t\right)  & -\Omega\left(  t\right)
\end{array}
\right)  \text{, and }\left[  \mathcal{U}\left(  t\right)  \right]
\overset{\text{def}}{=}\left(
\begin{array}
[c]{cc}%
\alpha\left(  t\right)  & \beta\left(  t\right) \\
-\beta^{\ast}\left(  t\right)  & \alpha^{\ast}\left(  t\right)
\end{array}
\right)  \text{,} \label{evol}%
\end{equation}
respectively. The unitarity of the quantum evolution demands that the complex
probability amplitudes $\alpha\left(  t\right)  $ and $\beta\left(  t\right)
$ fulfill the normalization condition, $\left\vert \alpha\left(  t\right)
\right\vert ^{2}+\left\vert \beta\left(  t\right)  \right\vert ^{2}=1$.
Once\textbf{ }the unitary evolution operator $\mathcal{U}\left(  t\right)  $
in Eq. (\ref{evol}) is given, the temporal evolution of a quantum source state
$\left\vert s\right\rangle $,%
\begin{equation}
\left\vert s\right\rangle \overset{\text{def}}{=}x\left\vert w\right\rangle
+\sqrt{1-x^{2}}\left\vert w_{\perp}\right\rangle \text{,} \label{evol1}%
\end{equation}
can be characterized by the following transformation law,%
\begin{equation}
\binom{x}{\sqrt{1-x^{2}}}\rightarrow\binom{\alpha\left(  t\right)
x+\beta\left(  t\right)  \sqrt{1-x^{2}}}{-\beta^{\ast}\left(  t\right)
x+\alpha^{\ast}\left(  t\right)  \sqrt{1-x^{2}}}\text{,} \label{evol2}%
\end{equation}
with $x\overset{\text{def}}{=}\left\langle w|s\right\rangle $ denoting the
quantum overlap. The set of orthonormal state vectors\textbf{ }$\left\{
\left\vert w\right\rangle \text{, }\left\vert w_{\perp}\right\rangle \right\}
$\textbf{ }generate the two-dimensional search\textbf{ }space\textbf{ }of
the\textbf{ }$N=2^{n}$-dimensional complex Hilbert space\textbf{ }%
$\mathcal{H}_{2}^{n}$. Therefore, using Eqs. (\ref{evol}), (\ref{evol1}), and
(\ref{evol2}), the probability that the source state $\left\vert
s\right\rangle $ transitions into the target state $\left\vert w\right\rangle
$ under $\mathcal{U}\left(  t\right)  $ becomes%
\begin{equation}
\mathcal{P}_{\left\vert s\right\rangle \rightarrow\left\vert w\right\rangle
}\left(  t\right)  \overset{\text{def}}{=}\left\vert \left\langle
w|\mathcal{U}\left(  t\right)  |s\right\rangle \right\vert ^{2}=\left\vert
\alpha\left(  t\right)  \right\vert ^{2}x^{2}+\left\vert \beta\left(
t\right)  \right\vert ^{2}\left(  1-x^{2}\right)  +\left[  \alpha\left(
t\right)  \beta^{\ast}\left(  t\right)  +\alpha^{\ast}\left(  t\right)
\beta\left(  t\right)  \right]  x\sqrt{1-x^{2}}\text{.} \label{good1}%
\end{equation}
It is evident from Eq. (\ref{good1}) that in order\textbf{\ }to calculate the
exact analytical expression of transition probabilities, one must\textbf{
}have the exact analytical formula of the evolution operator $\mathcal{U}%
\left(  t\right)  $ expressed in terms of the complex probability amplitudes
$\alpha\left(  t\right)  $ and $\beta\left(  t\right)  $. Ideally, having
specified the fields $\omega\left(  t\right)  $ and $\Omega\left(  t\right)  $
by means of physical arguments, one would solve the coupled system of first
order ordinary differential equations with time-dependent coefficients
emerging from the relation $i\hslash\mathcal{\dot{U}}\left(  t\right)
=\mathcal{H}_{\mathrm{su}\left(  2\text{; }%
\mathbb{C}
\right)  }\mathcal{U}\left(  t\right)  $ with $\mathcal{U}\left(  0\right)
=\mathcal{I}$,%
\begin{equation}
i\hslash\dot{\alpha}\left(  t\right)  =\Omega\left(  t\right)  \alpha\left(
t\right)  -\omega\left(  t\right)  \beta^{\ast}\left(  t\right)  \text{, and
}i\hslash\dot{\beta}\left(  t\right)  =\omega\left(  t\right)  \alpha^{\ast
}\left(  t\right)  +\Omega\left(  t\right)  \beta\left(  t\right)  \text{,}
\label{lode}%
\end{equation}
where $\alpha\left(  0\right)  =1$ and $\beta\left(  0\right)  =0$. However,
it is often the case that this general approach does not yield exact
analytical solutions. It is recognized that it is rather difficult to find
exact analytical solutions in time-dependent two-level quantum systems
specified by equations as the ones in Eq. (\ref{lode}). A very powerful
technique for investigating these types of quantum evolutions is the so-called
rotating coordinates technique. This method, originally proposed by Rabi,
Ramsey, and Schwinger in magnetic resonance problems\textbf{, }can be
presented as a three-step technique: first, recast\textbf{ }the original
problem from a stationary to a rotating frame of reference by performing a
suitable change of coordinates; second, find the solution to the simplified
problem viewed in a rotating frame of reference by means of rotating
coordinates; third, solve the original problem by an inverse transformation
from the rotating to the stationary frame. Usually, the rotating frame of
reference rotates about the axis specified by the magnetic field $\vec
{B}_{\parallel}$ with the angular frequency characterized by the magnetic
field $\vec{B}_{\perp}$. In quantum terms, assume that in a static frame of
reference the quantum evolution of the qubit is ruled\textbf{ }by the
Schr\"{o}dinger equation,%
\begin{equation}
i\hslash\partial_{t}\left\vert \psi\left(  t\right)  \right\rangle
=\mathcal{H}\left(  t\right)  \left\vert \psi\left(  t\right)  \right\rangle
\text{.}%
\end{equation}
Then, considering a unitary transformation $T$ between two distinct vector
bases $\left\{  \left\vert \psi\left(  t\right)  \right\rangle \right\}  $ and
$\left\{  \left\vert \psi^{\prime}\left(  t\right)  \right\rangle \right\}  $,
the Schr\"{o}dinger equation in the new frame of reference can be written as%
\begin{equation}
i\hslash\partial_{t}\left\vert \psi^{\prime}\left(  t\right)  \right\rangle
=\mathcal{H}^{\prime}\left(  t\right)  \left\vert \psi^{\prime}\left(
t\right)  \right\rangle \text{,}%
\end{equation}
with $\left\vert \psi^{\prime}\left(  t\right)  \right\rangle \overset
{\text{def}}{=}T\left\vert \psi\left(  t\right)  \right\rangle $ and
$\mathcal{H}^{\prime}\left(  t\right)  \overset{\text{def}}{=}\left[
T\mathcal{H}\left(  t\right)  T^{\dagger}+i\hslash\left(  \partial
_{t}T\right)  T^{\dagger}\right]  $. More specifically, assuming that the
unitary transformation $T\overset{\text{def}}{=}e^{-i\sigma_{z}\frac
{\phi_{\omega}\left(  t\right)  }{2}}$ is a rotation by an angle $\phi
_{\omega}\left(  t\right)  $ about the $z$-axis, the Hamiltonians in Eqs.
(\ref{easy2}) and (\ref{easy3}) can be recast as%
\begin{equation}
\mathcal{H}_{\mathrm{su}\left(  2\text{; }%
\mathbb{C}
\right)  }^{\prime}\left(  t\right)  \overset{\text{def}}{=}\left[
\Omega\left(  t\right)  +\frac{\hslash}{2}\dot{\phi}_{\omega}\left(  t\right)
\right]  \sigma_{z}+\omega_{\mathcal{H}}\left(  t\right)  \sigma_{x}\text{,}
\label{rabi1a}%
\end{equation}
and,%
\begin{equation}
\mathcal{H}_{\mathrm{su}\left(  2\text{; }%
\mathbb{C}
\right)  }^{\prime}\left(  t\right)  \overset{\text{def}}{=}\left[
-\frac{e\hslash}{2mc}B_{\parallel}\left(  t\right)  +\frac{\hslash}{2}%
\dot{\phi}_{\omega}\left(  t\right)  \right]  \sigma_{z}+\frac{e\hslash}%
{2mc}B_{\perp}\left(  t\right)  \sigma_{x}\text{,} \label{rabi2a}%
\end{equation}
respectively, where $\dot{\phi}_{\omega}\overset{\text{def}}{=}d\phi_{\omega
}/dt$. In the so-called original Rabi scenario, $\phi_{\omega}\left(
t\right)  \overset{\text{def}}{=}\omega_{0}t$ with the angular frequency
$\omega_{0}$ being a negative real constant while the magnetic field
intensities $B_{\parallel}$ and $B_{\perp}$ are constant. In particular,
assuming the so-called static resonance condition,%
\begin{equation}
\omega_{0}=\frac{e}{mc}B_{\parallel}\text{,}%
\end{equation}
the Hamiltonian in Eq. (\ref{rabi2a}) specifies a time-independent quantum
mechanical problem. Unlike the original Rabi scenario, in the so-called
generalized Rabi scenario proposed\textbf{ }by Messina and collaborators in
Refs. \cite{messina14,grimaudo18}, $B_{\parallel}$, $B_{\perp}$, and
$\phi_{\omega}$ are arbitrary time-dependent dynamical variables. In
particular, assuming the so-called generalized Rabi condition,%
\begin{equation}
\dot{\phi}_{\omega}\left(  t\right)  =-\frac{2}{\hslash}\Omega\left(
t\right)  \text{,} \label{GRCa}%
\end{equation}
the Hamiltonians in Eqs. (\ref{rabi1a}) and (\ref{rabi2a}) do not
yield\textbf{ }a time-independent quantum mechanical problem. Inspired by our
research reported in Ref. \cite{carloIJQI} and exploiting the findings in
Refs. \cite{messina14,grimaudo18}, we take into\textbf{ }consideration here
four quantum mechanical scenarios where the transition probability
$\mathcal{P}_{\left\vert w_{\perp}\right\rangle \rightarrow\left\vert
w\right\rangle }\left(  t\right)  $ from an initial state $\left\vert
w_{\perp}\right\rangle $ to a final state $\left\vert w\right\rangle $, with
$\left\langle w_{\perp}|w\right\rangle =\delta_{w_{\perp}\text{, }w}$,
$\sigma_{z}\left\vert w\right\rangle =+\left\vert w\right\rangle $, and
$\sigma_{z}\left\vert w_{\perp}\right\rangle =-\left\vert w_{\perp
}\right\rangle $, can be analytically described. \ In all four cases, we
assume to be in a physical scenario in which,%
\begin{equation}
\dot{\phi}_{\omega}\left(  t\right)  =\omega_{0}\text{, and }\Omega\left(
t\right)  =-\frac{\hslash}{2}\omega_{0}\text{,} \label{chosena}%
\end{equation}
with $\omega_{0}$ being\textbf{ }a negative real constant. We point out that,
from a pure mathematical viewpoint, more general temporal behaviors of
$\dot{\phi}_{\omega}\left(  t\right)  $ and $\Omega\left(  t\right)  $ in Eq.
(\ref{chosena}) could have been selected provided Eq. (\ref{GRCa}) is
fulfilled. However, the choice adopted in\ Eq. (\ref{chosena}) seems to be
more suitable from an experimental standpoint. The four scenarios can be
formally distinguished by means of the temporal behavior of the intensity
$\omega_{\mathcal{H}}\left(  t\right)  $ of the complex transverse
field\textbf{ }$\omega\left(  t\right)  $. In the first case, we assume a
constant field intensity $\omega_{\mathcal{H}}\left(  t\right)  $,%
\begin{equation}
\omega_{\mathcal{H}}^{\left(  1\right)  }\left(  t\right)  \overset
{\text{def}}{=}\Gamma\text{.} \label{32}%
\end{equation}
This first case specifies the original Rabi scenario where $\mathcal{P}%
_{\left\vert w_{\perp}\right\rangle \rightarrow\left\vert w\right\rangle
}\left(  t\right)  $ is given by,%
\begin{equation}
\mathcal{P}_{\left\vert w_{\perp}\right\rangle \rightarrow\left\vert
w\right\rangle }^{\left(  1\right)  }\left(  t\right)  =\sin^{2}\left(
\frac{\Gamma}{\hslash}t\right)  \text{.} \label{tp1}%
\end{equation}
In the remaining three cases, we take into consideration three generalized
Rabi scenarios where the field intensity $\omega_{\mathcal{H}}\left(
t\right)  $ manifests oscillatory, power law decay, and exponential law decay
behaviors,%
\begin{equation}
\omega_{\mathcal{H}}^{\left(  2\right)  }\left(  t\right)  \overset
{\text{def}}{=}\Gamma\cos\left(  \lambda t\right)  \text{, }\omega
_{\mathcal{H}}^{\left(  3\right)  }\left(  t\right)  \overset{\text{def}}%
{=}\frac{\Gamma}{\left(  1+\lambda t\right)  ^{2}}\text{, and }\omega
_{\mathcal{H}}^{\left(  4\right)  }\left(  t\right)  \overset{\text{def}}%
{=}\Gamma e^{-\lambda t}\text{,} \label{33}%
\end{equation}
respectively. Note that $\omega_{\mathcal{H}}^{\left(  2\right)  }\left(
t\right)  $ in Eq. (\ref{33})\ is a positive quantity on a temporal scale with
$0\leq t\leq\left(  \pi/2\right)  \lambda^{-1}$. In all three cases, it can be
proven that the transition probability $\mathcal{P}_{\left\vert w_{\perp
}\right\rangle \rightarrow\left\vert w\right\rangle }^{\left(  j\right)
}\left(  t\right)  $ is given by \cite{grimaudo18},%
\begin{equation}
\mathcal{P}_{\left\vert w_{\perp}\right\rangle \rightarrow\left\vert
w\right\rangle }^{\left(  j\right)  }\left(  t\right)  =\sin^{2}\left[
\int_{0}^{t}\frac{\omega_{\mathcal{H}}^{\left(  j\right)  }\left(  t^{\prime
}\right)  }{\hslash}dt^{\prime}\right]  \text{,} \label{tp2}%
\end{equation}
for any $j\in\left\{  2\text{, }3\text{, }4\right\}  $. Interestingly, being
on resonance, the transition probabilities In Eqs. (\ref{tp1}) and (\ref{tp2})
for all four cases depend only on the integral of the transverse field
intensity $\omega_{\mathcal{H}}\left(  t\right)  $.

Before beginning our information geometric analysis, we formally introduce
here an adimensional parameter $\beta_{0}$ that quantifies the departure from
the on-resonance condition,%
\begin{equation}
\beta_{0}\overset{\text{def}}{=}\frac{\hslash}{2\omega_{\mathcal{H}}\left(
t\right)  }\left[  \dot{\phi}_{\omega}\left(  t\right)  +\frac{2}{\hslash
}\Omega\left(  t\right)  \right]  \text{.} \label{betazero}%
\end{equation}
When $\beta_{0}$ assumes a nonzero constant value, the generalized expression
of the transition probability in Eq. (\ref{tp2}) is given by \cite{grimaudo18}%
,%
\begin{equation}
\mathcal{P}_{\left\vert w_{\perp}\right\rangle \rightarrow\left\vert
w\right\rangle }^{\left(  j\right)  }\left(  t\text{; }\beta_{0}\right)
\overset{\text{def}}{=}\frac{1}{1+\beta_{0}^{2}}\sin^{2}\left[  \sqrt
{1+\beta_{0}^{2}}\int_{0}^{t}\frac{\omega_{\mathcal{H}}^{\left(  j\right)
}\left(  t^{\prime}\right)  }{\hslash}dt^{\prime}\right]  \text{.} \label{tp3}%
\end{equation}
We emphasize that $\beta_{0}$ in Eq. (\ref{betazero}) is generally a
time-dependent quantity. However, in what follows, we shall limit our
information geometric analysis to the case in which $\beta_{0}$ is a constant
parameter with transition probabilities given in Eq. (\ref{tp3}). For further
details on the link between $\beta_{0}$ in Eq. (\ref{betazero}) and its
analogue counterpart in the framework of Rabi's original static off-resonance
condition, we refer to Appendix A.

\section{Information geometric analysis}

In this section, we present our information geometric analysis of
off-resonance effects on the chosen quantum driving strategies. For each
scheme, we analyze the consequences of a departure from the on-resonance
condition in terms of both geodesic paths and geodesic speeds on the
corresponding manifold of transition probability vectors.

\subsection{Preliminaries}

In order to provide some conceptual background for our forthcoming information
geometric analysis that emerges from quantum driving strategies, we begin by
introducing some preliminary remarks.

Assume to consider an $N\overset{\text{def}}{=}2^{n}$-dimensional complex
Hilbert space $\mathcal{H}_{2}^{n}$ together with two neighboring normalized
pure states $\left\vert \psi\left(  \theta\right)  \right\rangle $ and
$\left\vert \psi^{\prime}\left(  \theta\right)  \right\rangle $ where,
\begin{equation}
\left\vert \psi\right\rangle \overset{\text{def}}{=}\sum_{l=1}^{N}\sqrt{p_{l}%
}e^{i\varphi_{l}}\left\vert l\right\rangle \text{ and }\left\vert \psi
^{\prime}\right\rangle \overset{\text{def}}{=}\sum_{l=1}^{N}\sqrt{p_{l}%
+dp_{l}}e^{i\left(  \varphi_{l}+d\varphi_{l}\right)  }\left\vert
l\right\rangle \text{,}\label{explicit}%
\end{equation}
respectively. In Eq. (\ref{explicit}), $\left\{  \left\vert l\right\rangle
\right\}  $ with $1\leq l\leq N$ denotes an orthonormal basis of
$\mathcal{H}_{2}^{n}$ while $p_{l}=p_{l}\left(  \theta\right)  $ and
$\varphi_{l}=\varphi_{l}\left(  \theta\right)  $ are real functions of a
continuous real parameter $\theta$. The distinguishability metric on this
manifold of Hilbert space rays is specified by the Fubini-Study metric
\cite{caves94},%
\begin{equation}
ds_{\text{FS}}^{2}\overset{\text{def}}{=}\left\{  \cos^{-1}\left[  \left\vert
\left\langle \psi^{\prime}|\psi\right\rangle \right\vert \right]  \right\}
^{2}=g_{ab}^{\left(  \text{FS}\right)  }\left(  \theta\right)  d\theta
^{a}d\theta^{b}\text{,}\label{fubini}%
\end{equation}
where the Fubini-Study metric tensor components $g_{ab}^{\left(
\text{FS}\right)  }$ are related to the Fisher-Rao metric tensor components
$\mathcal{F}_{ab}\left(  \theta\right)  $ by the condition\textbf{,}%
\begin{equation}
g_{ab}^{\left(  \text{FS}\right)  }\left(  \theta\right)  =\frac{1}{4}\left[
\mathcal{F}_{ab}\left(  \theta\right)  +4\sigma_{ab}^{2}\left(  \theta\right)
\right]  \text{.}\label{fubini2}%
\end{equation}
The quantities $\mathcal{F}_{ab}\left(  \theta\right)  $ and $\sigma_{ab}%
^{2}\left(  \theta\right)  $ in Eq. (\ref{fubini2}) are given by,%
\begin{equation}
\mathcal{F}_{ab}\left(  \theta\right)  \overset{\text{def}}{=}\sum_{l=1}%
^{N}\frac{1}{p_{l}\left(  \theta\right)  }\frac{\partial p_{l}\left(
\theta\right)  }{\partial\theta^{a}}\frac{\partial p_{l}\left(  \theta\right)
}{\partial\theta^{b}}\text{,}\label{FS12}%
\end{equation}
and,%
\begin{equation}
\sigma_{ab}^{2}\left(  \theta\right)  \overset{\text{def}}{=}\sum_{l=1}%
^{N}\frac{\partial\varphi_{l}\left(  \theta\right)  }{\partial\theta^{a}}%
\frac{\partial\varphi_{l}\left(  \theta\right)  }{\partial\theta^{b}}%
p_{l}\left(  \theta\right)  -\left(  \sum_{k=1}^{N}\frac{\partial\varphi
_{l}\left(  \theta\right)  }{\partial\theta^{a}}p_{l}\left(  \theta\right)
\right)  \left(  \sum_{k=1}^{N}\frac{\partial\varphi_{l}\left(  \theta\right)
}{\partial\theta^{b}}p_{l}\left(  \theta\right)  \right)  \text{,}%
\label{variance}%
\end{equation}
respectively. In what follows, we assume that the non-negative term
$\sigma_{ab}^{2}\left(  \theta\right)  $ in Eq. (\ref{variance})
\textbf{(}which denotes the variance of the phase changes\textbf{)} is equal
to zero. This working assumption can be justified by rephasing in a suitable
fashion the basis vectors used to write the state $\left\vert \psi\left(
\theta\right)  \right\rangle $\textbf{. }Specifically, the rephasing procedure
demands that $\operatorname{Im}\left[  \left\langle \psi\left(  \theta\right)
|l\right\rangle \left\langle l|d\psi_{\perp}\left(  \theta\right)
\right\rangle \right]  =0$, for any $1\leq l\leq N$. The state $\left\vert
d\psi_{\perp}\right\rangle \overset{\text{def}}{=}\left\vert d\psi
\right\rangle -\left\langle \psi|d\psi\right\rangle \left\vert d\psi
\right\rangle $ denotes the projection of $\left\vert d\psi\right\rangle $
orthogonal to $\left\vert \psi\right\rangle $ where $\left\vert d\psi
\right\rangle \overset{\text{def}}{=}\left\vert \psi^{\prime}\right\rangle
-\left\vert \psi\right\rangle $ while $\left\vert \psi\right\rangle $ and
$\left\vert \psi^{\prime}\right\rangle $ are given in Eq. (\ref{explicit}). In
summary, for a convenient choice of the basis vectors $\left\{  \left\vert
l\right\rangle \right\}  $ used for the decomposition in Eq. (\ref{explicit}),
$g_{ab}^{\left(  \text{FS}\right)  }\left(  \theta\right)  $ becomes
proportional to $\mathcal{F}_{ab}\left(  \theta\right)  $ as evident from Eq.
(\ref{fubini2}). We refer to Ref. \cite{caves94} for further details. For the
sake of completeness, we emphasize that if the basis $\left\{  \left\vert
l\right\rangle \right\}  $ satisfies the above mentioned conditions, the basis
vectors can always be rephased so that both $\left\langle l|\psi\right\rangle
$ and $\left\langle l|d\psi_{\perp}\right\rangle $ are real. These latter
conditions, in turn, are reminiscent of the so-called parallel transport
conditions that occur in the context of Berry's description of a geometric
effect in the shape of an additional phase factor emerging after an adiabatic
and cyclic transport of a quantum system \cite{berry84}. Specifically, in the
framework of Berry's phase analysis generalized by Aharonov and Anandan by
removing the adiabaticity constraint in the cyclic quantum evolution
\cite{anandan87}, it is possible to show that there exists one distinct curve
in the Hilbert space $\mathcal{H}$ fulfilling the parallel transport
conditions. Namely, two neighboring states $\left\vert \psi\left(
\theta\right)  \right\rangle $ and $\left\vert \psi\left(  \theta
+d\theta\right)  \right\rangle $ in $\mathcal{H}$ have the same phase, that is
to say, $\left\langle \psi\left(  \theta\right)  |\psi\left(  \theta
+d\theta\right)  \right\rangle $ is real and positive. For such a curve, the
dynamical phase vanishes. In summary, for special Hamiltonians that satisfy a
special gauge choice (that is, the parallel transport conditions), the
dynamical phase vanishes \cite{resta00}.

In our paper, the parameter $\theta$ is the statistical version of the elapsed
time $t$\textbf{. }That is, we assume $\theta$ is a parameter that can be
experimentally specified by measurement of a suitable observable that changes
with time (for instance, the transverse magnetic field intensity\textbf{
}$B_{\perp}\left(  t\right)  $). \ For further details on the concept of
statistical elapsed time, we refer to Ref. \cite{brau996}. In particular, we
suppose that the output $\left\vert \psi\left(  \theta\right)  \right\rangle $
of a quantum driving Hamiltonian acting on the input defined by the normalized
source state $\left\vert s\right\rangle \overset{\text{def}}{=}\left\vert
\psi\left(  \theta_{0}\right)  \right\rangle $ can be described as,%
\begin{equation}
\left\vert \psi\left(  \theta_{0}\right)  \right\rangle \mapsto\left\vert
\psi\left(  \theta\right)  \right\rangle \overset{\text{def}}{=}%
e^{i\varphi_{w}\left(  \theta\right)  }\sqrt{p_{w}\left(  \theta\right)
}\left\vert w\right\rangle +e^{i\varphi_{w_{\perp}}\left(  \theta\right)
}\sqrt{p_{w_{\perp}}\left(  \theta\right)  }\left\vert w_{\perp}\right\rangle
\text{.} \label{output}%
\end{equation}
In general, the normalized output state $\left\vert \psi\left(  \theta\right)
\right\rangle $ is an element of the two-dimensional subspace of the $n$-qubit
complex Hilbert space $\mathcal{H}_{2}^{n}$ accommodating the source state
$\left\vert s\right\rangle $ and generated by the set of orthonormal state
vectors $\left\{  \left\vert w\right\rangle \text{, }\left\vert w_{\perp
}\right\rangle \right\}  $. The modulus squared of the probability amplitudes
$p_{w}\left(  \theta\right)  \overset{\text{def}}{=}\left\vert \left\langle
w|\psi\left(  \theta\right)  \right\rangle \right\vert ^{2}$ and $p_{w_{\perp
}}\left(  \theta\right)  \overset{\text{def}}{=}\left\vert \left\langle
w_{\perp}|\psi\left(  \theta\right)  \right\rangle \right\vert ^{2}$ stand for
the success and failure probabilities of the given driving strategy,
respectively. Furthermore, $\varphi_{w}\left(  \theta\right)  $ and
$\varphi_{w_{\perp}}\left(  \theta\right)  $ denote the real quantum phases of
the states $\left\vert w\right\rangle $ and $\left\vert w_{\perp}\right\rangle
$, respectively. We note that the quantum state $\left\vert \psi\left(
\theta\right)  \right\rangle $ in\ Eq. (\ref{output}) is parameterized by
means of a single continuous real parameter that emerges from the elapsed time
of the driving strategy. As previously mentioned, this parameter $\theta$
mimics a statistical macrovariable used to distinguish neighboring quantum
states $\left\vert \psi\left(  \theta\right)  \right\rangle $ and $\left\vert
\psi\left(  \theta\right)  \right\rangle +\left\vert d\psi\left(
\theta\right)  \right\rangle $ along a path through the space of pure quantum
states. In summary, we focus on the space of probability distributions
$\left\{  p\left(  \theta\right)  \right\}  $,%
\begin{equation}
\left\vert \psi\left(  \theta\right)  \right\rangle \mapsto p\left(
\theta\right)  \overset{\text{def}}{=}\left(  p_{w}\left(  \theta\right)
\text{, }p_{w_{\perp}}\left(  \theta\right)  \right)  =\left(  \left\vert
\left\langle w|\psi\left(  \theta\right)  \right\rangle \right\vert
^{2}\text{, }\left\vert \left\langle w_{\perp}|\psi\left(  \theta\right)
\right\rangle \right\vert ^{2}\right)  \text{,} \label{output2}%
\end{equation}
where the natural Riemannian distinguishability metric between two neighboring
probability distributions $p\left(  \theta\right)  $ and $p\left(
\theta+d\theta\right)  $ is given by the Fisher-Rao information metric in Eq.
(\ref{FS12}). For an overview of the use of information geometric techniques
to analog quantum search problems, we refer to Refs.
\cite{carlopre,cafaroaip,cafaro12,cafaro17}.

In any quantum driving scheme considered in this paper, following Refs.
\cite{messina14,grimaudo18}\textbf{,} we assume the success probability
$p_{0}\left(  \theta\right)  $ and the failure probability $p_{1}\left(
\theta\right)  $ can be recast as%
\begin{equation}
p_{0}\left(  \theta\right)  \overset{\text{def}}{=}\mathcal{A}\sin^{2}\left[
\Sigma\left(  \theta\right)  \right]  \text{ and, }p_{1}\left(  \theta\right)
\overset{\text{def}}{=}1-\mathcal{A}\sin^{2}\left[  \Sigma\left(
\theta\right)  \right]  \text{,} \label{probability}%
\end{equation}
respectively. In Eq. (\ref{probability}), the quantity $\theta$ denotes the
temporal parameter while the quantity $\mathcal{A}$ is defined as%
\begin{equation}
\mathcal{A}=\mathcal{A}\left(  \beta_{0}\right)  \overset{\text{def}}{=}%
\frac{1}{1+\beta_{0}^{2}}\text{,} \label{a}%
\end{equation}
with $\beta_{0}$ being the (assumed) constant parameter that quantifies the
deviation from the on-resonance condition. The quantity $\Sigma\left(
\theta\right)  $, instead, specifies the type of driving being considered. The
Fisher information $\mathcal{F}\left(  \theta\right)  $ corresponding to the
probability vector $\left(  p_{0}\left(  \theta\right)  \text{, }p_{1}\left(
\theta\right)  \right)  $ is given by \cite{amari,felice18},%
\begin{equation}
\mathcal{F}\left(  \theta\right)  \overset{\text{def}}{=}\frac{\dot{p}_{0}%
^{2}\left(  \theta\right)  }{p_{0}\left(  \theta\right)  }+\frac{\dot{p}%
_{1}^{2}\left(  \theta\right)  }{p_{1}\left(  \theta\right)  }\text{,}
\label{fisher}%
\end{equation}
with $\dot{p}\overset{\text{def}}{=}dp/d\theta$. Substituting Eq.
(\ref{probability}) into Eq. (\ref{fisher}), the Fisher information
$\mathcal{F}\left(  \theta\right)  $ becomes%
\begin{equation}
\mathcal{F}\left(  \theta\right)  \overset{\text{def}}{=}4\mathcal{A}%
^{2}\left(  \frac{d\Sigma}{d\theta}\right)  ^{2}\left\{  \frac{\cos^{2}\left[
\Sigma\left(  \theta\right)  \right]  }{\mathcal{A}}+\frac{\cos^{2}\left[
\Sigma\left(  \theta\right)  \right]  \sin^{2}\left[  \Sigma\left(
\theta\right)  \right]  }{1-\mathcal{A}\sin^{2}\left[  \Sigma\left(
\theta\right)  \right]  }\right\}  \text{.} \label{fisherg}%
\end{equation}
Note that when we are on-resonance, $\beta_{0}=0$, $\mathcal{A}=1$, and
$\mathcal{F}\left(  \theta\right)  $ reduces to%
\begin{equation}
\mathcal{F}\left(  \theta\right)  \overset{\text{def}}{=}4\left(
\frac{d\Sigma}{d\theta}\right)  ^{2}\text{.}%
\end{equation}
In particular, when $\Sigma\left(  \theta\right)  $ is linear in $\theta$,
$\mathcal{F}\left(  \theta\right)  =4$ as reported in Ref. \cite{carlopre}.
Finally, once the expression of the Fisher information $\mathcal{F}\left(
\theta\right)  $ is given, the geodesic equation to integrate becomes%
\begin{equation}
\frac{d^{2}\theta}{d\xi^{2}}+\frac{1}{2\mathcal{F}\left(  \theta\right)
}\frac{d\mathcal{F}\left(  \theta\right)  }{d\theta}\left(  \frac{d\theta
}{d\xi}\right)  ^{2}=0\text{.} \label{geo1a}%
\end{equation}
For more details on how to derive Eq. (\ref{geo1a}) and how to express the
quantum mechanical infinitesimal Fubini-Study line element in terms of the
Fisher information and/or the Fisher-Rao information metric, we refer to Ref.
\cite{carlopre}.\textbf{ }For the sake of completeness, we observe that Eq.
(\ref{geo1a}) can be recast as,%
\begin{equation}
\ddot{\theta}+\frac{\mathcal{\dot{F}}}{2\mathcal{F}}\dot{\theta}=0\text{,}
\label{geo2a}%
\end{equation}
where $\dot{\theta}\overset{\text{def}}{=}d\theta/d\xi$ and $\mathcal{\dot{F}%
}\overset{\text{def}}{=}d\mathcal{F}/d\xi$. Defining $w\overset{\text{def}}%
{=}\dot{\theta}$ and after some straightforward algebraic manipulations, Eq.
(\ref{geo2a}) reduces to%
\begin{equation}
\frac{d\mathcal{F}}{\mathcal{F}}=-2\frac{dw}{w}\text{.} \label{geo3a}%
\end{equation}
Integration of Eq. (\ref{geo3a}) yields a closed-form implicit relation
between the dependent variable $\theta$ and the independent variable $\xi$,%
\begin{equation}
c\left(  \theta\text{, }\xi\right)  =\int_{\theta\left(  \xi_{0}\right)
}^{\theta\left(  \xi\right)  }\sqrt{\mathcal{F}\left(  \theta\right)  }%
d\theta-\int_{\xi_{0}}^{\xi}a_{0}d\xi=0\text{,} \label{constraint}%
\end{equation}
with $a_{0}$ being a \emph{real} constant of integration. Eq.
(\ref{constraint}) is an implicit constraint equation between the variables
$\theta$ and $\xi$. Despite the closed-form implicit relation in Eq.
(\ref{constraint}), in what follows we shall numerically find the geodesic
paths satisfying Eq. (\ref{geo1a}) for the four driving strategies defined in
Eqs. (\ref{32}) and (\ref{33}).

\subsection{Geodesic paths}

In what follows, after specifying $\Sigma\left(  \theta\right)  $ in Eq.
(\ref{probability}) and $\mathcal{F}\left(  \theta\right)  $ in Eq.
(\ref{fisherg}), we shall focus on numerically finding from Eq. (\ref{geo1a})
the geodesic paths $\theta=\theta\left(  \xi\right)  $ that correspond to the
various off-resonant driving schemes.

\subsubsection{Constant behavior}

In the first case , we assume $\Sigma\left(  \theta\right)  $ is defined as
\begin{equation}
\Sigma\left(  \theta\right)  \overset{\text{def}}{=}\mathcal{B}\theta\text{.}
\label{sigma1}%
\end{equation}
Therefore, from Eqs. (\ref{probability}) and (\ref{sigma1}), the probabilities
$p_{0}\left(  \theta\right)  $ and $p_{1}\left(  \theta\right)  $ are given by%
\begin{equation}
p_{0}\left(  \theta\right)  \overset{\text{def}}{=}\mathcal{A}\sin^{2}\left[
\mathcal{B}\theta\right]  \text{ and, }p_{1}\left(  \theta\right)
\overset{\text{def}}{=}1-\mathcal{A}\sin^{2}\left[  \mathcal{B}\theta\right]
\text{,} \label{probability1}%
\end{equation}
respectively. In Eq. (\ref{probability1}), the quantity $\mathcal{A}$ is
defined in Eq. (\ref{a}) while $\mathcal{B}$ is given by%
\begin{equation}
\mathcal{B}=\mathcal{B}\left(  \beta_{0}\text{, }\Gamma\text{, }%
\hslash\right)  \overset{\text{def}}{=}\sqrt{1+\beta_{0}^{2}}\frac{\Gamma
}{\hslash}\text{,} \label{b}%
\end{equation}
respectively. In Fig. $1$, we plot the success probability $p_{0}\left(
\theta\right)  $ versus the temporal parameter $\theta$ in the case of
constant driving with the assumption that $\beta_{0}\in\left\{  0\text{,
}1/4\text{, }1/2\text{, }1\right\}  $. From Fig. $1$, we note that the maximum
value of the success probability decreases with respect to increasing values
of the parameter $\beta_{0}$. Furthermore, the periodic oscillatory behavior
of the success probability is also affected by nonzero values of $\beta_{0}$.
In particular, to higher values of $\beta_{0}$ there correspond smaller
(higher) values of the period (frequency) of oscillation. In this first case,
using Eq. (\ref{probability1}) into Eq. (\ref{fisherg}), the Fisher
information becomes%
\begin{equation}
\mathcal{F}_{\text{constant}}^{\left(  \text{off-resonance}\right)  }\left(
\theta\right)  \overset{\text{def}}{=}4\mathcal{AB}^{2}\left\{  \frac
{1+\cos\left(  2\mathcal{B}\theta\right)  }{2-\mathcal{A}\left[  1-\cos\left(
2\mathcal{B}\theta\right)  \right]  }\right\}  \text{.} \label{fisher1}%
\end{equation}
For the sake of completeness, note that when $\beta_{0}=0$, $\mathcal{A}=1$,
and $\mathcal{F}\left(  \theta\right)  $ in Eq. (\ref{fisher1}) reduces to%
\begin{equation}
\mathcal{F}_{\text{constant}}^{\left(  \text{on-resonance}\right)  }\left(
\theta\right)  =4\mathcal{B}^{2}\text{.} \label{fisher1a}%
\end{equation}
Finally, employing Eqs. (\ref{geo1a}) and (\ref{fisher1}), the geodesic
equation to integrate becomes%
\begin{equation}
\frac{d^{2}\theta}{d\xi^{2}}-\frac{\left(  1-\mathcal{A}\right)
\mathcal{B}\tan\left(  \mathcal{B}\theta\right)  }{1-\mathcal{A}\sin
^{2}\left(  \mathcal{B}\theta\right)  }\left(  \frac{d\theta}{d\xi}\right)
^{2}=0\text{.} \label{cumpa1}%
\end{equation}
The output arising from the numerical integration of Eq. (\ref{cumpa1}) is
plotted in Fig. $2$.

\begin{figure}[t]
\label{fig1}\centering
\includegraphics[width=0.35\textwidth] {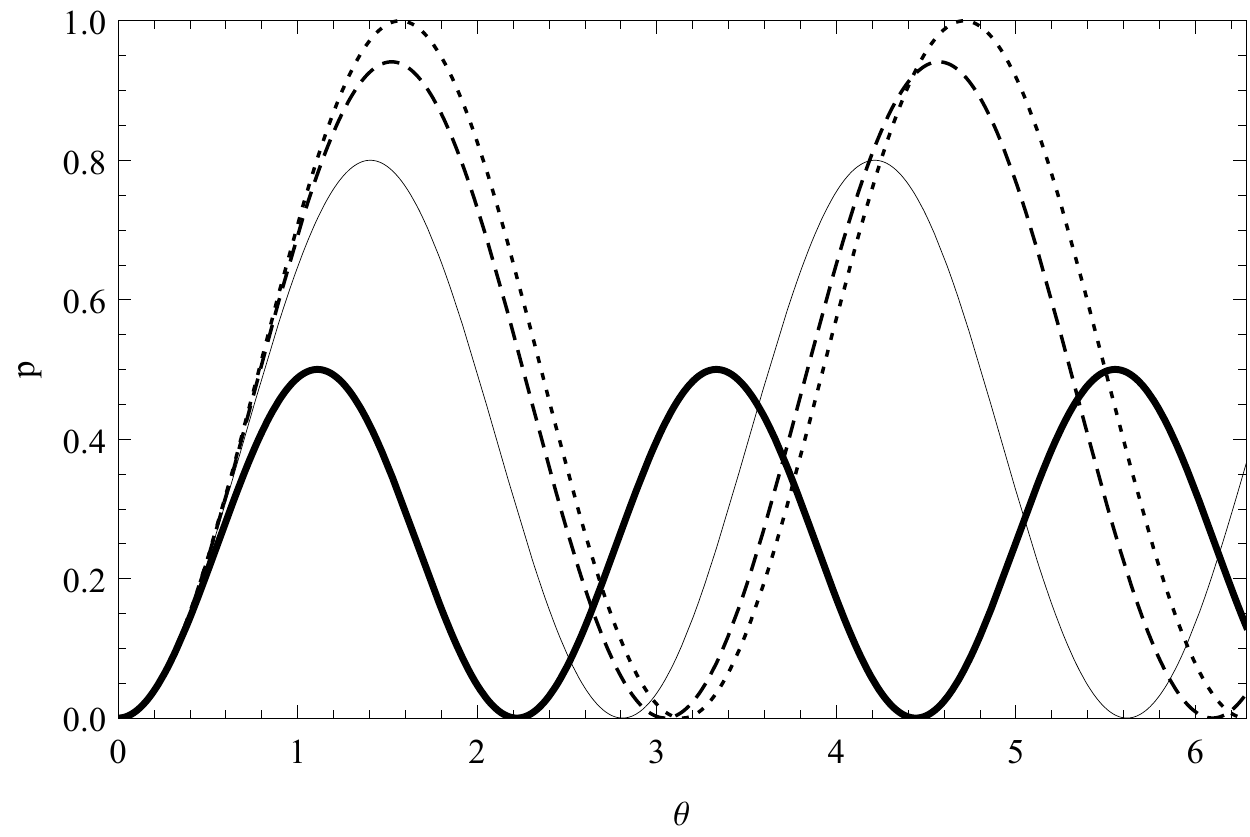}\caption{Plot of the success
probability $p$ versus the temporal parameter $\theta$ in the case of constant
driving for $\beta=0$ (dotted), $\beta_{0}=0.25$ (dashed), $\beta_{0}=0.5$
(thin solid), and $\beta_{0}=1$ (thick solid).}%
\end{figure}

\subsubsection{Oscillatory behavior}

In this second scenario, we assume $\Sigma\left(  \theta\right)  $ is given
by
\begin{equation}
\Sigma\left(  \theta\right)  \overset{\text{def}}{=}\frac{\mathcal{B}}%
{\lambda}\sin\left(  \lambda\theta\right)  \text{.} \label{sigma2}%
\end{equation}
Therefore, from Eqs. (\ref{probability}) and (\ref{sigma2}), the probabilities
$p_{0}\left(  \theta\right)  $ and $p_{1}\left(  \theta\right)  $ become%
\begin{equation}
p_{0}\left(  \theta\right)  \overset{\text{def}}{=}\mathcal{A}\sin^{2}\left[
\frac{\mathcal{B}}{\lambda}\sin\left(  \lambda\theta\right)  \right]  \text{
and, }p_{1}\left(  \theta\right)  \overset{\text{def}}{=}1-\mathcal{A}\sin
^{2}\left[  \frac{\mathcal{B}}{\lambda}\sin\left(  \lambda\theta\right)
\right]  \text{,}%
\end{equation}
respectively. In this case, employing Eq. (\ref{sigma2}), the Fisher
information in Eq. (\ref{fisherg}) becomes
\begin{equation}
\mathcal{F}_{\text{oscillatory}}^{\left(  \text{off-resonance}\right)
}\left(  \theta\right)  \overset{\text{def}}{=}\frac{\left\{  \frac{d\left(
\mathcal{A}\sin^{2}\left[  \frac{\mathcal{B}}{\lambda}\sin\left(
\lambda\theta\right)  \right]  \right)  }{d\theta}\right\}  ^{2}}%
{\mathcal{A}\sin^{2}\left[  \frac{\mathcal{B}}{\lambda}\sin\left(
\lambda\theta\right)  \right]  }+\frac{\left\{  \frac{d\left(  1-\mathcal{A}%
\sin^{2}\left[  \frac{\mathcal{B}}{\lambda}\sin\left(  \lambda\theta\right)
\right]  \right)  }{dx}\right\}  ^{2}}{1-\mathcal{A}\sin^{2}\left[
\frac{\mathcal{B}}{\lambda}\sin\left(  \lambda\theta\right)  \right]
}\text{,} \label{fisher2}%
\end{equation}
that is, after some straightforward but tedious algebra,%
\begin{equation}
\mathcal{F}_{\text{oscillatory}}^{\left(  \text{off-resonance}\right)
}\left(  \theta\right)  =8\mathcal{AB}^{2}\cos^{2}\left(  \lambda
\theta\right)  \frac{\cos^{2}\left[  \frac{\mathcal{B}}{\lambda}\sin\left(
\lambda\theta\right)  \right]  }{2-\mathcal{A}\left\{  1-\cos\left[
2\frac{\mathcal{B}}{\lambda}\sin\left(  \lambda\theta\right)  \right]
\right\}  }\text{.} \label{fisher2A}%
\end{equation}
Observe that when $\beta_{0}=0$, $\mathcal{A}=1$, and, recalling that
$1+\cos\left(  2x\right)  =2\cos^{2}\left(  x\right)  $, $\mathcal{F}\left(
\theta\right)  $ in Eq. (\ref{fisher2A}) reduces to%
\begin{equation}
\mathcal{F}_{\text{oscillatory}}^{\left(  \text{on-resonance}\right)  }\left(
\theta\right)  =4\mathcal{B}^{2}\cos^{2}(\lambda\theta)\text{.}
\label{fisher2a}%
\end{equation}
Finally, employing Eqs. (\ref{geo1a})\textbf{ }and (\ref{fisher2A}), the
geodesic equation to integrate in this second scenario can be formally written
as%
\begin{equation}
\frac{d^{2}\theta}{d\xi^{2}}+\frac{1}{2\mathcal{F}_{\text{o}}\left(
\theta\right)  }\frac{d\mathcal{F}_{\text{o}}\left(  \theta\right)  }{d\theta
}\left(  \frac{d\theta}{d\xi}\right)  ^{2}=0\text{, } \label{w1}%
\end{equation}
with $\mathcal{F}_{\text{o}}\left(  \theta\right)  \overset{\text{def}}%
{=}\mathcal{F}_{\text{oscillatory}}^{\left(  \text{off-resonance}\right)
}\left(  \theta\right)  $ as in Eq. (\ref{fisher2A}). Using a symbolic
mathematical software (Mathematica, for instance) together with the analytical
knowledge of what happens in the limiting case of $\beta_{0}=0$
\cite{carlotechnical19}, Eq. (\ref{w1}) can be finally recast as
\begin{equation}
\frac{d^{2}\theta}{d\xi^{2}}-\left\{  \lambda\tan\left(  \lambda\theta\right)
+\frac{\left(  1-\mathcal{A}\right)  \mathcal{B}\cos\left(  \lambda
\theta\right)  \tan\left[  \frac{\mathcal{B}}{\lambda}\sin\left(
\lambda\theta\right)  \right]  }{1-\mathcal{A}\sin^{2}\left[  \frac
{\mathcal{B}}{\lambda}\sin\left(  \lambda\theta\right)  \right]  }\right\}
\left(  \frac{d\theta}{d\xi}\right)  ^{2}=0\text{.} \label{cumpa2}%
\end{equation}
The output emerging from numerically integrating Eq. (\ref{cumpa2}) is plotted
in Fig. $2$.

\subsubsection{Power law decay}

In this third scenario, we assume $\Sigma\left(  \theta\right)  $ is defined
as
\begin{equation}
\Sigma\left(  \theta\right)  \overset{\text{def}}{=}\frac{\mathcal{B}}%
{\lambda}\left(  1-\frac{1}{1+\lambda\theta}\right)  \text{.} \label{sigma3}%
\end{equation}
Therefore, the probabilities $p_{0}\left(  \theta\right)  $ and $p_{1}\left(
\theta\right)  $ in Eq. (\ref{probability}) can be rewritten as%
\begin{equation}
p_{0}\left(  \theta\right)  \overset{\text{def}}{=}\mathcal{A}\sin^{2}\left[
\frac{\mathcal{B}}{\lambda}\left(  1-\frac{1}{1+\lambda\theta}\right)
\right]  \text{ and, }p_{1}\left(  \theta\right)  \overset{\text{def}}%
{=}1-\mathcal{A}\sin^{2}\left[  \frac{\mathcal{B}}{\lambda}\left(  1-\frac
{1}{1+\lambda\theta}\right)  \right]  \text{,}%
\end{equation}
respectively. In this case, the Fisher information in Eq. (\ref{fisherg}) can
be formally recast as%
\begin{equation}
\mathcal{F}_{\text{power-law-decay}}^{\left(  \text{off-resonance}\right)
}\left(  \theta\right)  \overset{\text{def}}{=}\frac{\left\{  \frac{d\left(
\mathcal{A}\sin^{2}\left[  \frac{\mathcal{B}}{\lambda}\left(  1-\frac
{1}{1+\lambda\theta}\right)  \right]  \right)  }{d\theta}\right\}  ^{2}%
}{\mathcal{A}\sin^{2}\left[  \frac{\mathcal{B}}{\lambda}\left(  1-\frac
{1}{1+\lambda\theta}\right)  \right]  }+\frac{\left\{  \frac{d\left(
1-\mathcal{A}\sin^{2}\left[  \frac{\mathcal{B}}{\lambda}\left(  1-\frac
{1}{1+\lambda\theta}\right)  \right]  \right)  }{dx}\right\}  ^{2}%
}{1-\mathcal{A}\sin^{2}\left[  \frac{\mathcal{B}}{\lambda}\left(  1-\frac
{1}{1+\lambda\theta}\right)  \right]  }\text{,} \label{fisher3}%
\end{equation}
that is, after some tedious algebra,%
\begin{equation}
\mathcal{F}_{\text{power-law-decay}}^{\left(  \text{off-resonance}\right)
}\left(  \theta\right)  =\frac{4\mathcal{AB}^{2}}{\left(  1+\lambda
\theta\right)  ^{4}}\frac{1+\cos\left(  2\frac{\mathcal{B}\theta}%
{1+\lambda\theta}\right)  }{2-\mathcal{A}\left[  1-\cos\left(  2\frac
{\mathcal{B}\theta}{1+\lambda\theta}\right)  \right]  }\text{.}
\label{fisher3B}%
\end{equation}
Once again, for the sake of completeness, we point out that when $\beta_{0}%
=0$, $\mathcal{A}=1$, and $\mathcal{F}\left(  \theta\right)  $ in Eq.
(\ref{fisher3B}) reduces to%
\begin{equation}
\mathcal{F}_{\text{power-law-decay}}^{\left(  \text{on-resonance}\right)
}\left(  \theta\right)  =\frac{4\mathcal{B}^{2}}{\left(  1+\lambda
\theta\right)  ^{4}}\text{.} \label{fisher3a}%
\end{equation}
Finally,\textbf{ }making use of Eqs. (\ref{geo1a})\textbf{ }and
(\ref{fisher3B}), the geodesic equation to integrate in this third scenario
can be formally written as%
\begin{equation}
\frac{d^{2}\theta}{d\xi^{2}}+\frac{1}{2\mathcal{F}_{\text{pld}}\left(
\theta\right)  }\frac{d\mathcal{F}_{\text{pld}}\left(  \theta\right)
}{d\theta}\left(  \frac{d\theta}{d\xi}\right)  ^{2}=0\text{, } \label{w2}%
\end{equation}
with $\mathcal{F}_{\text{pld}}\left(  \theta\right)  \overset{\text{def}}%
{=}\mathcal{F}_{\text{power-law-decay}}^{\left(  \text{off-resonance}\right)
}\left(  \theta\right)  $ as in Eq. (\ref{fisher3B}). Employing a symbolic
mathematical software (Mathematica, for instance) together with the analytical
knowledge of what occurs in the on-resonance case where $\beta_{0}=0$
\cite{carlotechnical19}, Eq. (\ref{w2}) can be finally recast as%
\begin{equation}
\frac{d^{2}\theta}{d\xi^{2}}-\left\{  \frac{2\lambda}{1+\lambda\theta}%
+\frac{(1-\mathcal{A})\mathcal{B}\tan\left(  \frac{\mathcal{B}\theta
}{1+\lambda\theta}\right)  }{\left(  1+\lambda\theta\right)  ^{2}\left[
1-\mathcal{A}\sin^{2}\left(  \frac{\mathcal{B}\theta}{1+\lambda\theta}\right)
\right]  }\right\}  \left(  \frac{d\theta}{d\xi}\right)  ^{2}=0\text{.}
\label{cumpa3}%
\end{equation}
In Fig. $2$, we plot the output that arises from the numerical integration of
Eq. (\ref{cumpa3}).

\subsubsection{Exponential decay}

In this last scenario, we assume%
\begin{equation}
\Sigma\left(  \theta\right)  \overset{\text{def}}{=}\frac{\mathcal{B}}%
{\lambda}\left(  1-e^{-\lambda\theta}\right)  \text{.}%
\end{equation}
Therefore, the success and failure probabilities probabilities $p_{0}\left(
\theta\right)  $ and $p_{1}\left(  \theta\right)  $ are given by%
\begin{equation}
p_{0}\left(  \theta\right)  \overset{\text{def}}{=}\mathcal{A}\sin^{2}\left[
\frac{\mathcal{B}}{\lambda}\left(  1-e^{-\lambda\theta}\right)  \right]
\text{ and, }p_{1}\left(  \theta\right)  \overset{\text{def}}{=}%
1-\mathcal{A}\sin^{2}\left[  \frac{\mathcal{B}}{\lambda}\left(  1-e^{-\lambda
\theta}\right)  \right]  \text{,}%
\end{equation}
respectively. In this case, the Fisher information in Eq. (\ref{fisherg}) is
formally given by%
\begin{equation}
\mathcal{F}_{\text{exponential-decay}}^{\left(  \text{off-resonance}\right)
}\left(  \theta\right)  \overset{\text{def}}{=}\frac{\left\{  \frac{d\left(
\mathcal{A}\sin^{2}\left[  \frac{\mathcal{B}}{\lambda}\left(  1-e^{-\lambda
\theta}\right)  \right]  \right)  }{d\theta}\right\}  ^{2}}{\mathcal{A}%
\sin^{2}\left[  \frac{\mathcal{B}}{\lambda}\left(  1-e^{-\lambda\theta
}\right)  \right]  }+\frac{\left\{  \frac{d\left(  1-\mathcal{A}\sin
^{2}\left[  \frac{\mathcal{B}}{\lambda}\left(  1-e^{-\lambda\theta}\right)
\right]  \right)  }{dx}\right\}  ^{2}}{1-\mathcal{A}\sin^{2}\left[
\frac{\mathcal{B}}{\lambda}\left(  1-e^{-\lambda\theta}\right)  \right]
}\text{,}%
\end{equation}
that is,%
\begin{equation}
\mathcal{F}_{\text{exponential-decay}}^{\left(  \text{off-resonance}\right)
}\left(  \theta\right)  \overset{\text{def}}{=}4\mathcal{AB}^{2}%
e^{-2\lambda\theta}\frac{1+\cos\left[  \frac{2\mathcal{B}}{\lambda}\left(
1-e^{-\lambda\theta}\right)  \right]  }{2-\mathcal{A}\left\{  1-\cos\left[
\frac{2\mathcal{B}}{\lambda}\left(  1-e^{-\lambda\theta}\right)  \right]
\right\}  } \label{fisher4B}%
\end{equation}
Observe that when $\beta_{0}=0$, $\mathcal{A}=1$, $\mathcal{F}\left(
\theta\right)  $ in Eq. (\ref{fisher4B}) reduces to%
\begin{equation}
\mathcal{F}_{\text{exponential-decay}}^{\left(  \text{on-resonance}\right)
}\left(  \theta\right)  =4\mathcal{B}^{2}e^{-2\lambda\theta}\text{.}
\label{fisher4a}%
\end{equation}
Lastly, using Eqs. (\ref{geo1a}) and (\ref{fisher4B}), the geodesic equation
to integrate in this third scenario can be formally written as%
\begin{equation}
\frac{d^{2}\theta}{d\xi^{2}}+\frac{1}{2\mathcal{F}_{\text{exp}}\left(
\theta\right)  }\frac{d\mathcal{F}_{\text{exp}}\left(  \theta\right)
}{d\theta}\left(  \frac{d\theta}{d\xi}\right)  ^{2}=0\text{, } \label{w3}%
\end{equation}
with $\mathcal{F}_{\text{exp}}\left(  \theta\right)  \overset{\text{def}}%
{=}\mathcal{F}_{\text{exponential-decay}}^{\left(  \text{off-resonance}%
\right)  }\left(  \theta\right)  $ as in Eq. (\ref{fisher4B}). Making use of a
symbolic mathematical software (Mathematica, for instance) together with the
analytical knowledge of what happens in the limiting case of $\beta_{0}=0$
\cite{carlotechnical19}, Eq. (\ref{w3}) can be finally recast as%
\begin{equation}
\frac{d^{2}\theta}{d\xi^{2}}-\left\{  \lambda+\frac{\left(  1-\mathcal{A}%
\right)  \mathcal{B}e^{-\lambda\theta}\tan\left[  \frac{\mathcal{B}}{\lambda
}\left(  1-e^{-\lambda\theta}\right)  \right]  }{1-\mathcal{A}\sin^{2}\left[
\frac{\mathcal{B}}{\lambda}\left(  1-e^{-\lambda\theta}\right)  \right]
}\right\}  \left(  \frac{d\theta}{d\xi}\right)  ^{2}=0\text{.} \label{cumpa4}%
\end{equation}
In Fig. $2$, we report the numerical plots of the geodesic paths $\theta$
versus the affine parameter $\xi$ for the four driving strategies with
geodesic equations in\ Eqs. (\ref{cumpa1}), (\ref{cumpa2}), (\ref{cumpa3}),
and (\ref{cumpa4}). In the LHS of Fig. $2$, we consider the on-resonance
scenario with $\beta_{0}=0$. In the RHS of Fig. $2$, instead, we consider the
off-resonance scenario with $\beta_{0}=1/2$. In both sides, we set
$\Gamma/\hslash=1$, $\lambda=2/\pi$, $\theta_{0}=0$, and $\dot{\theta}_{0}=1$.
The most noticeable feature in Fig. $2$ is the deviation from the straight
line behavior of the geodesic path that emerges from the constant driving
scheme when the on-resonance condition is satisfied. In particular, a first
visual comparison of the corresponding plots in the two cases (on-resonance
v.s. off-resonance) seems to suggest that the constant driving scheme might be
the least robust to departures from the on-resonance condition. Indeed, this
preliminary remark will be quantitatively confirmed in our forthcoming
information geometric analysis.

\begin{figure}[t]
\label{fig2}\centering
\includegraphics[width=0.75\textwidth]{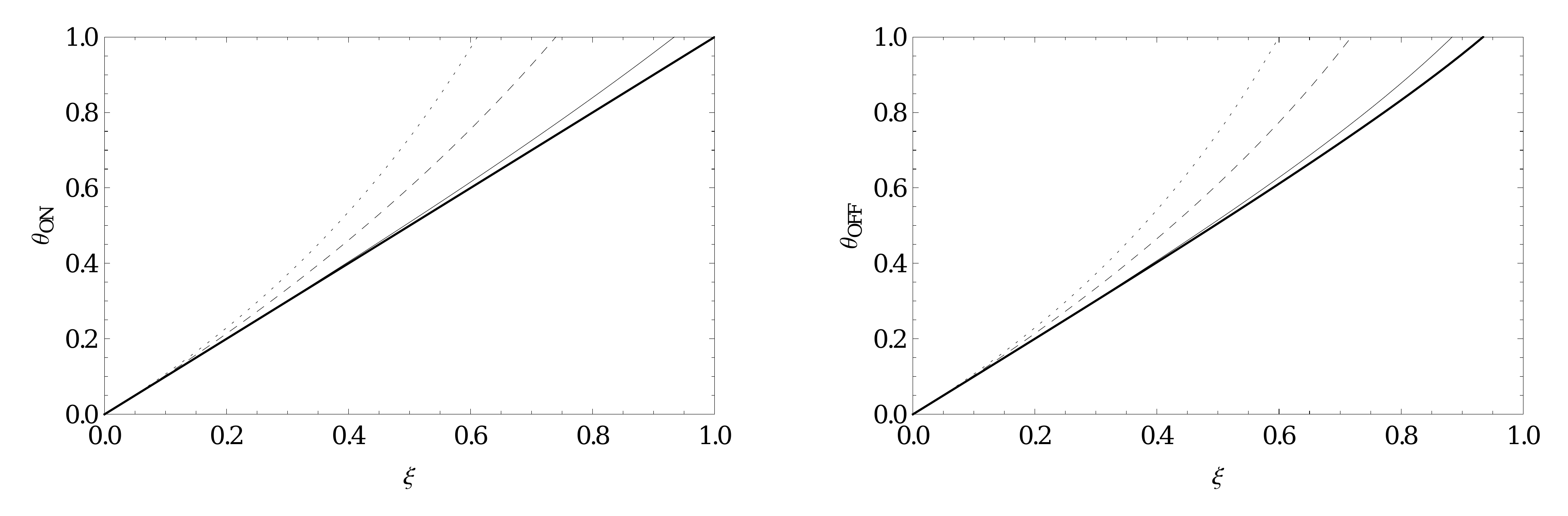}\caption{Numerical plots of the
geodesic paths $\theta$ versus the affine parameter $\xi$ for the four driving
strategies: Power-law (dotted line), exponential (dashed line), oscillatory
(thin solid line), and constant (thick solid line). In the LHS
(left-hand-side), we consider the on-resonance scenario with $\beta_{0}=0$. In
the RHS (right-hand-side), we consider the off-resonance scenario with
$\beta_{0}=1/2$. In both sides, we set $\Gamma/\hslash=1$, $\lambda=2/\pi$,
$\theta_{0}=0$, and $\dot{\theta}_{0}=1$.}%
\end{figure}

\subsection{Geodesic speeds}

In what follows, after specifying $\Sigma\left(  \theta\right)  $ in Eq.
(\ref{probability}) and $\mathcal{F}\left(  \theta\right)  $ in Eq.
(\ref{fisherg}), we shall focus on finding the geodesic speeds $v$,%
\begin{equation}
v\overset{\text{def}}{=}\frac{1}{2}\sqrt{\mathcal{F}\left(  \theta\right)
}\frac{d\theta}{d\xi}\text{,} \label{vgeo}%
\end{equation}
corresponding to the various off-resonant driving schemes. For some technical
details on the proof of the fact that geodesic paths have constant speed, we
refer to Appendix B.

\subsubsection{Constant case}

In this first case, using Eqs. (\ref{a}), (\ref{b}), and (\ref{fisher1}), the
on-resonance geodesic speed $v$ from Eq. (\ref{vgeo}) is given by%
\begin{equation}
v_{\text{ON}}=\frac{\Gamma}{\hslash}\dot{\theta}\text{.} \label{geo1}%
\end{equation}
In the off-resonance regime, instead, the geodesic speed $v$ from Eq.
(\ref{vgeo}) can be recast as%
\begin{equation}
v_{\text{OFF}}=\frac{\Gamma}{\hslash}\left\{  \frac{1+\cos\left(
2\frac{\Gamma}{\hslash}\sqrt{1+\beta_{0}^{2}}\theta\right)  }{2-\frac
{1}{1+\beta_{0}^{2}}\left[  1-\cos\left(  2\frac{\Gamma}{\hslash}\sqrt
{1+\beta_{0}^{2}}\theta\right)  \right]  }\right\}  ^{\frac{1}{2}}\dot{\theta
}\text{.} \label{geo2}%
\end{equation}
Furthermore, setting $\Gamma/\hslash=1$, $\dot{\theta}\left(  \xi_{0}\right)
=\dot{\theta}_{0}=1$, and $\theta\left(  \xi_{0}\right)  =\theta_{0}$,
$v_{\text{ON}}$ and $v_{\text{OFF}}$ in Eqs. (\ref{geo1}) and (\ref{geo2})
become%
\begin{equation}
v_{\text{ON}}^{\left(  \text{constant}\right)  }\overset{\text{def}}%
{=}1\text{,} \label{geo3}%
\end{equation}
and,%
\begin{equation}
v_{\text{OFF}}^{\left(  \text{constant}\right)  }\left(  \beta_{0}\text{,
}\theta_{0}\right)  \overset{\text{def}}{=}\left\{  \frac{1+\cos\left(
2\sqrt{1+\beta_{0}^{2}}\theta_{0}\right)  }{2-\frac{1}{1+\beta_{0}^{2}}\left[
1-\cos\left(  2\sqrt{1+\beta_{0}^{2}}\theta_{0}\right)  \right]  }\right\}
^{\frac{1}{2}}\text{,} \label{geo4}%
\end{equation}
respectively. From Eqs. (\ref{geo3}) and (\ref{geo4}), we note that the
geodesic speed becomes sensitive to the initial condition $\theta_{0}$ when
departing from the on-resonance scenario in the case of a driving strategy
with constant complex transverse field intensity\textbf{.} Finally, employing
Eqs. (\ref{geo4}) and (\ref{geo3}), we define the robustness coefficient as
the ratio%
\begin{equation}
r_{\text{constant}}\left(  \beta_{0}\text{, }\theta_{0}\right)  \overset
{\text{def}}{=}\frac{v_{\text{OFF}}^{\left(  \text{constant}\right)  }\left(
\beta_{0}\text{, }\theta_{0}\right)  }{v_{\text{ON}}^{\left(  \text{constant}%
\right)  }\left(  \theta_{0}\right)  }\text{,} \label{r1}%
\end{equation}
that is,%
\begin{equation}
r_{\text{constant}}\left(  \beta_{0}\text{, }\theta_{0}\right)  =\left\{
\frac{1+\cos\left(  2\sqrt{1+\beta_{0}^{2}}\theta_{0}\right)  }{2-\frac
{1}{1+\beta_{0}^{2}}\left[  1-\cos\left(  2\sqrt{1+\beta_{0}^{2}}\theta
_{0}\right)  \right]  }\right\}  ^{\frac{1}{2}}\text{.}%
\end{equation}
In Fig. $3$, we plot the analytical (thin solid) and numerical (filled circle)
values of the geodesic speed $v_{\text{OFF}}$ in Eq. (\ref{geo2}) versus the
initial condition $\theta_{0}$ in the case of constant driving. In the plot,
we set $\beta_{0}=1/2$, $\Gamma/\hslash=1$, and $\dot{\theta}_{0}=1$. The plot
in Fig. $3$ clearly illustrates the emergence of the sensitivity on the
initial condition $\theta_{0}$ of the geodesic speed $v_{\text{OFF}}$ for the
constant driving scheme when the on-resonance condition is violated. For the
sake of completeness, we point out that despite the fact that this sensitivity
is present in the remaining three driving strategies even when the
on-resonance condition is fulfilled, off-resonance effects substantially
enhance this sensitivity as evident from our information geometric analysis of
geodesic speeds presented here.

\begin{figure}[t]
\label{fig3}\centering
\includegraphics[width=0.35\textwidth]{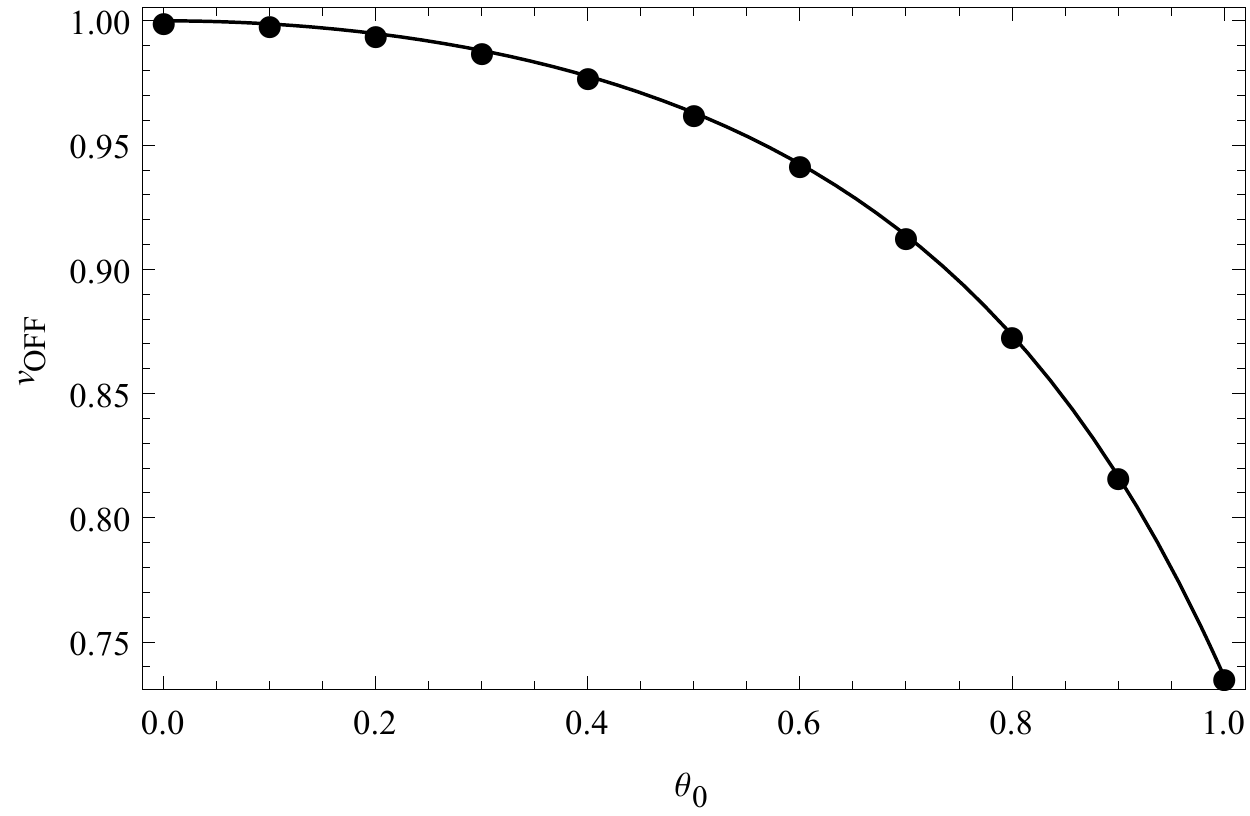}\caption{Plot of the analytical
(thin solid) and numerical (filled circle) values of the geodesic speed
$v_{\text{OFF}}$ versus the initial condition $\theta_{0}$ in the case of
constant driving. In the plot, we set $\beta_{0}=1/2$, $\Gamma/\hslash=1$, and
$\dot{\theta}_{0}=1$.}%
\end{figure}

\subsubsection{Oscillatory behavior}

In this second case, making use of Eqs. (\ref{a}), (\ref{b}), and
(\ref{fisher2A}), the on-resonance geodesic speed $v$ from Eq. (\ref{vgeo})
becomes%
\begin{equation}
v_{\text{ON}}=\frac{\Gamma}{\hslash}\left\vert \cos\left(  \lambda
\theta\right)  \right\vert \dot{\theta}\text{.} \label{geo5}%
\end{equation}
In the off-resonance regime, instead, the geodesic speed $v_{\text{OFF}}$ is
given by%
\begin{equation}
v_{\text{OFF}}=\frac{\Gamma}{\hslash}\left\vert \cos\left(  \lambda
\theta\right)  \right\vert \left\{  \frac{2\cos^{2}\left[  \frac{1}{\lambda
}\frac{\Gamma}{\hslash}\sqrt{1+\beta_{0}^{2}}\sin\left(  \lambda\theta\right)
\right]  }{2-\frac{1}{1+\beta_{0}^{2}}\left\{  1-\cos\left[  \frac{2}{\lambda
}\frac{\Gamma}{\hslash}\sqrt{1+\beta_{0}^{2}}\sin\left(  \lambda\theta\right)
\right]  \right\}  }\right\}  ^{\frac{1}{2}}\dot{\theta}\text{.} \label{geo6}%
\end{equation}
Moreover, letting $\Gamma/\hslash=1$, $\lambda=2/\pi$, $\dot{\theta}\left(
\xi_{0}\right)  =\dot{\theta}_{0}=1$, and $\theta\left(  \xi_{0}\right)
=\theta_{0}$, $v_{\text{ON}}$ and $v_{\text{OFF}}$ in Eqs. (\ref{geo5}) and
(\ref{geo6}) become%
\begin{equation}
v_{\text{ON}}^{\left(  \text{oscillatory}\right)  }\overset{\text{def}}%
{=}\left\vert \cos\left(  \frac{2}{\pi}\theta_{0}\right)  \right\vert \text{,}
\label{geo7}%
\end{equation}
and,%
\begin{equation}
v_{\text{OFF}}^{\left(  \text{oscillatory}\right)  }\left(  \beta_{0}\text{,
}\theta_{0}\right)  \overset{\text{def}}{=}\left\vert \cos\left(  \frac{2}%
{\pi}\theta_{0}\right)  \right\vert \left\{  \frac{2\cos^{2}\left[  \frac{\pi
}{2}\sqrt{1+\beta_{0}^{2}}\sin\left(  \frac{2}{\pi}\theta_{0}\right)  \right]
}{2-\frac{1}{1+\beta_{0}^{2}}\left\{  1-\cos\left[  \pi\sqrt{1+\beta_{0}^{2}%
}\sin\left(  \frac{2}{\pi}\theta_{0}\right)  \right]  \right\}  }\right\}
^{\frac{1}{2}}\text{,} \label{geo8}%
\end{equation}
respectively. Finally, using Eqs. (\ref{geo7}) and (\ref{geo8}), we introduce
the robustness coefficient as the ratio%
\begin{equation}
r_{\text{oscillatory}}\left(  \beta_{0}\text{, }\theta_{0}\right)
\overset{\text{def}}{=}\frac{v_{\text{OFF}}^{\left(  \text{oscillatory}%
\right)  }\left(  \beta_{0}\text{, }\theta_{0}\right)  }{v_{\text{ON}%
}^{\left(  \text{oscillatory}\right)  }\left(  \theta_{0}\right)  }\text{,}
\label{r2}%
\end{equation}
that is,%
\begin{equation}
r_{\text{oscillatory}}\left(  \beta_{0}\text{, }\theta_{0}\right)  =\left\{
\frac{2\cos^{2}\left[  \frac{\pi}{2}\sqrt{1+\beta_{0}^{2}}\sin\left(  \frac
{2}{\pi}\theta_{0}\right)  \right]  }{2-\frac{1}{1+\beta_{0}^{2}}\left\{
1-\cos\left[  \pi\sqrt{1+\beta_{0}^{2}}\sin\left(  \frac{2}{\pi}\theta
_{0}\right)  \right]  \right\}  }\right\}  ^{\frac{1}{2}}\text{.}%
\end{equation}
From Eqs. (\ref{geo7}) and (\ref{geo8}), we observe that $v_{\text{OFF}%
}^{\left(  \text{oscillatory}\right)  }\left(  \beta_{0}\text{, }\theta
_{0}\right)  \leq v_{\text{ON}}^{\left(  \text{oscillatory}\right)  }$ and
$0\leq r_{\text{oscillatory}}\left(  \beta_{0}\text{, }\theta_{0}\right)  <1$.

\subsubsection{Power law decay}

In this third case, Eqs. (\ref{a}), (\ref{b}), and (\ref{fisher3B}) lead to an
expression of the on-resonance geodesic speed $v$ from Eq. (\ref{vgeo}) given
by%
\begin{equation}
v_{\text{ON}}=\frac{\Gamma}{\hslash}\frac{1}{\left(  1+\lambda\theta\right)
^{2}}\dot{\theta}\text{.} \label{geo9}%
\end{equation}
In the off-resonance regime where $\beta_{0}\neq0$, instead, the geodesic
speed $v_{\text{OFF}}$ from Eq. (\ref{vgeo}) becomes%
\begin{equation}
v_{\text{OFF}}=\frac{\Gamma}{\hslash}\frac{1}{\left(  1+\lambda\theta\right)
^{2}}\left\{  \frac{1+\cos\left(  \frac{2\frac{\Gamma}{\hslash}\sqrt
{1+\beta_{0}^{2}}\theta}{1+\lambda\theta}\right)  }{2-\frac{1}{1+\beta_{0}%
^{2}}\left[  1-\cos\left(  \frac{2\frac{\Gamma}{\hslash}\sqrt{1+\beta_{0}^{2}%
}\theta}{1+\lambda\theta}\right)  \right]  }\right\}  ^{\frac{1}{2}}%
\dot{\theta}\text{.} \label{geo10}%
\end{equation}
In addition, assuming $\Gamma/\hbar=1$, $\lambda=2/\pi$, $\dot{\theta}\left(
\xi_{0}\right)  =\dot{\theta}_{0}=1$, and $\theta\left(  \xi_{0}\right)
=\theta_{0}$, $v_{\text{ON}}$ and $v_{\text{OFF}}$ in Eqs. (\ref{geo9}) and
(\ref{geo10}) become%
\begin{equation}
v_{\text{ON}}^{\left(  \text{power-law-decay}\right)  }\overset{\text{def}}%
{=}\frac{1}{\left(  1+\frac{2}{\pi}\theta_{0}\right)  ^{2}}\text{,}
\label{geo11}%
\end{equation}
and,%
\begin{equation}
v_{\text{OFF}}^{\left(  \text{power-law-decay}\right)  }\left(  \beta
_{0}\text{, }\theta_{0}\right)  \overset{\text{def}}{=}\frac{1}{\left(
1+\frac{2}{\pi}\theta_{0}\right)  ^{2}}\left\{  \frac{1+\cos\left(
\frac{2\sqrt{1+\beta_{0}^{2}}\theta_{0}}{1+\frac{2}{\pi}\theta_{0}}\right)
}{2-\frac{1}{1+\beta_{0}^{2}}\left[  1-\cos\left(  \frac{2\sqrt{1+\beta
_{0}^{2}}\theta_{0}}{1+\frac{2}{\pi}\theta_{0}}\right)  \right]  }\right\}
^{\frac{1}{2}}\text{,} \label{geo12}%
\end{equation}
respectively. Lastly, employing Eqs. (\ref{geo11}) and (\ref{geo12}), we
define the robustness coefficient as the ratio%
\begin{equation}
r_{\text{power-law-decay}}\left(  \beta_{0}\text{, }\theta_{0}\right)
\overset{\text{def}}{=}\frac{v_{\text{OFF}}^{\left(  \text{power-law-decay}%
\right)  }\left(  \beta_{0}\text{, }\theta_{0}\right)  }{v_{\text{ON}%
}^{\left(  \text{power-law-decay}\right)  }\left(  \theta_{0}\right)
}\text{,} \label{r3}%
\end{equation}
that is,%
\begin{equation}
r_{\text{power-law-decay}}\left(  \beta_{0}\text{, }\theta_{0}\right)
=\left\{  \frac{1+\cos\left(  \frac{2\sqrt{1+\beta_{0}^{2}}\theta_{0}}%
{1+\frac{2}{\pi}\theta_{0}}\right)  }{2-\frac{1}{1+\beta_{0}^{2}}\left[
1-\cos\left(  \frac{2\sqrt{1+\beta_{0}^{2}}\theta_{0}}{1+\frac{2}{\pi}%
\theta_{0}}\right)  \right]  }\right\}  ^{\frac{1}{2}}\text{.}%
\end{equation}
From Eqs. (\ref{geo11}) and (\ref{geo12}), we observe that $v_{\text{OFF}%
}^{\left(  \text{power-law-decay}\right)  }\left(  \beta_{0}\text{, }%
\theta_{0}\right)  \leq v_{\text{ON}}^{\left(  \text{power-law-decay}\right)
}$ and\textbf{ }$0\leq r_{\text{power-law-decay}}\left(  \beta_{0}\text{,
}\theta_{0}\right)  <1$.

\begin{figure}[t]
\label{fig4}\centering
\includegraphics[width=1\textwidth]{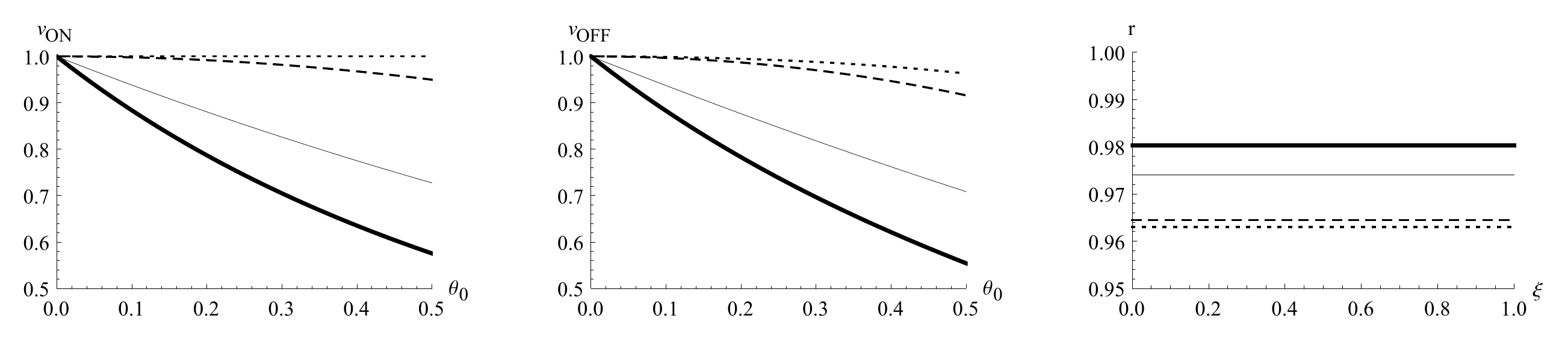}\caption{From left to right: (L) Plot
of the geodesic speed $v_{\text{ON}}$ versus the initial condition $\theta
_{0}$ with $\beta_{0}=0$; (C) Plot of the geodesic speed $v_{\text{OFF}}$
versus the initial condition $\theta_{0}$ with $\beta_{0}=1/2$; (R) Plot of
the robustness coefficient $r$ versus the affine parameter $\xi$. Plots that
correspond to the constant, oscillatory, exponential decay, and power law
decay strategies appear in dotted, dashed, thin solid, and thick solid lines,
respectively. In all plots, we set $\Gamma/\hslash=1$, $\lambda=2/\pi$, and
$\dot{\theta}_{0}=1$.}%
\end{figure}

\subsubsection{Exponential decay}

In this fourth and last case, making use of Eqs. (\ref{a}), (\ref{b}), and
(\ref{fisher4B}), the on-resonance geodesic speed $v$ from Eq. (\ref{vgeo})
becomes%
\begin{equation}
v_{\text{ON}}=\frac{\Gamma}{\hslash}e^{-\lambda\theta}\dot{\theta}\text{.}
\label{geo13}%
\end{equation}
In the off-resonance regime, instead, the geodesic speed $v_{\text{OFF}}$ is
given by%
\begin{equation}
v_{\text{OFF}}=\frac{\Gamma}{\hslash}e^{-\lambda\theta}\left\{  \frac
{1+\cos\left[  \frac{2}{\lambda}\frac{\Gamma}{\hslash}\sqrt{1+\beta_{0}^{2}%
}\left(  1-e^{-\lambda\theta}\right)  \right]  }{2-\frac{1}{1+\beta_{0}^{2}%
}\left\{  1-\cos\left[  \frac{2}{\lambda}\frac{\Gamma}{\hslash}\sqrt
{1+\beta_{0}^{2}}\left(  1-e^{-\lambda\theta}\right)  \right]  \right\}
}\right\}  ^{\frac{1}{2}}\dot{\theta}\text{.} \label{geo14}%
\end{equation}
Assuming $\Gamma/\hslash=1$, $\lambda=2/\hslash$, $\dot{\theta}\left(  \xi
_{0}\right)  =\dot{\theta}_{0}=1$, and $\theta\left(  \xi_{0}\right)
=\theta_{0}$, $v_{\text{ON}}$ and $v_{\text{OFF}}$ in Eqs. (\ref{geo13}) and
(\ref{geo14}) become%
\begin{equation}
v_{\text{ON}}^{\left(  \text{exponential-decay}\right)  }\overset{\text{def}%
}{=}e^{-\frac{2}{\pi}\theta_{0}}\text{,} \label{geo15}%
\end{equation}
and,%
\begin{equation}
v_{\text{OFF}}^{\left(  \text{exponential-decay}\right)  }\left(  \beta
_{0}\text{, }\theta_{0}\right)  \overset{\text{def}}{=}e^{-\frac{2}{\pi}%
\theta_{0}}\left\{  \frac{1+\cos\left[  \pi\sqrt{1+\beta_{0}^{2}}\left(
1-e^{-\frac{2}{\pi}\theta_{0}}\right)  \right]  }{2-\frac{1}{1+\beta_{0}^{2}%
}\left\{  1-\cos\left[  \pi\sqrt{1+\beta_{0}^{2}}\left(  1-e^{-\frac{2}{\pi
}\theta_{0}}\right)  \right]  \right\}  }\right\}  ^{\frac{1}{2}}\text{,}
\label{geo16}%
\end{equation}
respectively. Finally, by means of Eqs. (\ref{geo15}) and (\ref{geo16}), we
define the robustness coefficient as the ratio%
\begin{equation}
r_{\text{exponential-decay}}\left(  \beta_{0}\text{, }\theta_{0}\right)
\overset{\text{def}}{=}\frac{v_{\text{OFF}}^{\left(  \text{exponential-decay}%
\right)  }\left(  \beta_{0}\text{, }\theta_{0}\right)  }{v_{\text{ON}%
}^{\left(  \text{exponential-decay}\right)  }\left(  \theta_{0}\right)
}\text{,} \label{r4}%
\end{equation}
that is,%
\begin{equation}
r_{\text{exponential-decay}}\left(  \beta_{0}\text{, }\theta_{0}\right)
=\left\{  \frac{1+\cos\left[  \pi\sqrt{1+\beta_{0}^{2}}\left(  1-e^{-\frac
{2}{\pi}\theta_{0}}\right)  \right]  }{2-\frac{1}{1+\beta_{0}^{2}}\left\{
1-\cos\left[  \pi\sqrt{1+\beta_{0}^{2}}\left(  1-e^{-\frac{2}{\pi}\theta_{0}%
}\right)  \right]  \right\}  }\right\}  ^{\frac{1}{2}}\text{.}%
\end{equation}
In Fig. $4$, going from left to right, we plot: (L) Plot of the geodesic
speeds $v_{\text{ON}}$ in Eqs. (\ref{geo1}), (\ref{geo5}), (\ref{geo9}), and
(\ref{geo13}) versus the initial condition $\theta_{0}$ with $\beta_{0}=0$;
(C) Plot of the geodesic speeds $v_{\text{OFF}}$ in Eqs. (\ref{geo2}),
(\ref{geo6}), (\ref{geo10}), and (\ref{geo14}) versus the initial condition
$\theta_{0}$ with $\beta_{0}=1/2$; (R) Plot of the robustness coefficients $r$
in Eqs. (\ref{r1}), (\ref{r2}), (\ref{r3}), and (\ref{r4}) versus the affine
parameter $\xi$. Plots that correspond to the constant, oscillatory,
exponential decay, and power law decay driving strategies appear in dotted,
dashed, thin solid, and thick solid lines, respectively. In all plots, we set
$\Gamma/\hslash=1$, $\lambda=2/\pi$, and $\dot{\theta}_{0}=1$. The plots in
Fig. $4$ help exhibiting the effects of the off-resonance condition on the
speed of geodesic paths that correspond to the variety of driving strategies
being considered in this paper. In particular, we observe that to faster
(slower) driving strategies, there seem to correspond smaller (larger)
robustness coefficients. In other words, slower driving strategies appear to
be more robust against departures from the on-resonance condition. The ranking
of the various driving strategies is quite straightforward in the on-resonance
condition. However, when departing from this condition, off-resonance effects
lead to a richer set of dynamical scenarios. This, in turn, makes the
comparison of the performance of our chosen driving schemes more delicate. For
example, despite our illustrative depiction in Fig. $4$, it is possible to
uncover two-dimensional parametric regions\textbf{ }$\mathcal{P}\left(
\beta_{0}\text{, }\theta_{0}\right)  $\textbf{ }specified by the parameters
$\beta_{0}$ and $\theta_{0}$ where the constant driving scheme is not only the
least robust but also the slowest one. Indeed, we plot in Fig. $5$ a
two-dimensional parametric region $\mathcal{P}\left(  \beta_{0}\text{, }%
\theta_{0}\right)  $,%
\begin{equation}
\mathcal{P}\left(  \beta_{0}\text{, }\theta_{0}\right)  \overset{\text{def}%
}{=}\left\{  \left(  \beta_{0}\text{, }\theta_{0}\right)  \in%
\mathbb{R}
_{+}\times%
\mathbb{R}
_{+}:v_{\text{OFF}}^{\left(  \text{strategy-}i\right)  }\geq v_{\text{OFF}%
}^{\left(  \text{strategy-}j\right)  }\text{ and, }r^{\left(  \text{strategy-}%
i\right)  }\geq r^{\left(  \text{strategy-}j\right)  }\right\}  \text{,}%
\end{equation}
with $i\neq j\in\left\{  1\text{, }2\text{, }3\text{, }4\right\}  $ and where
the oscillatory (light grey), the exponential decay (grey), and the power law
decay (black) driving strategies outperform the constant driving scheme in
terms of both geodesic speed and robustness. More specifically, in Fig. $5$ we
have\textbf{ }%
\begin{equation}
\mathcal{P}_{\text{power-law-decay}}\left(  \beta_{0}\text{, }\theta
_{0}\right)  \subseteq\mathcal{P}_{\text{exponential-decay}}\left(  \beta
_{0}\text{, }\theta_{0}\right)  \subseteq\mathcal{P}_{\text{oscillatory}%
}\left(  \beta_{0}\text{, }\theta_{0}\right)  \text{,} \label{TG1}%
\end{equation}
and, in addition,%
\begin{equation}
r_{\text{power-law-decay}}\left(  \beta_{0}\text{, }\theta_{0}\right)  \geq
r_{\text{exponential-decay}}\left(  \beta_{0}\text{, }\theta_{0}\right)  \geq
r_{\text{oscillatory}}\left(  \beta_{0}\text{, }\theta_{0}\right)  \text{.}
\label{TG2}%
\end{equation}
In all region plots, we set $\Gamma/\hslash=1$, $\lambda=2/\pi$, and
$\dot{\theta}_{0}=1$. In addition, the maximum success probability of each
driving scheme is assumed to be greater than $25/26$, that is, $0\leq\beta
_{0}\lesssim0.20$ in Fig. $5$. The main take-home message from Fig. $5$\ is
that in the off-resonance scenario it is possible to uncover two-dimensional
parametric regions\textbf{ }$\mathcal{P}\left(  \beta_{0}\text{, }\theta
_{0}\right)  $\textbf{\ }where the best on-resonance driving scheme (that is,
the constant one) can be outperformed both in terms of speed and
robustness\textbf{.}

\begin{figure}[t]
\label{fig5}\centering
\includegraphics[width=0.35\textwidth]{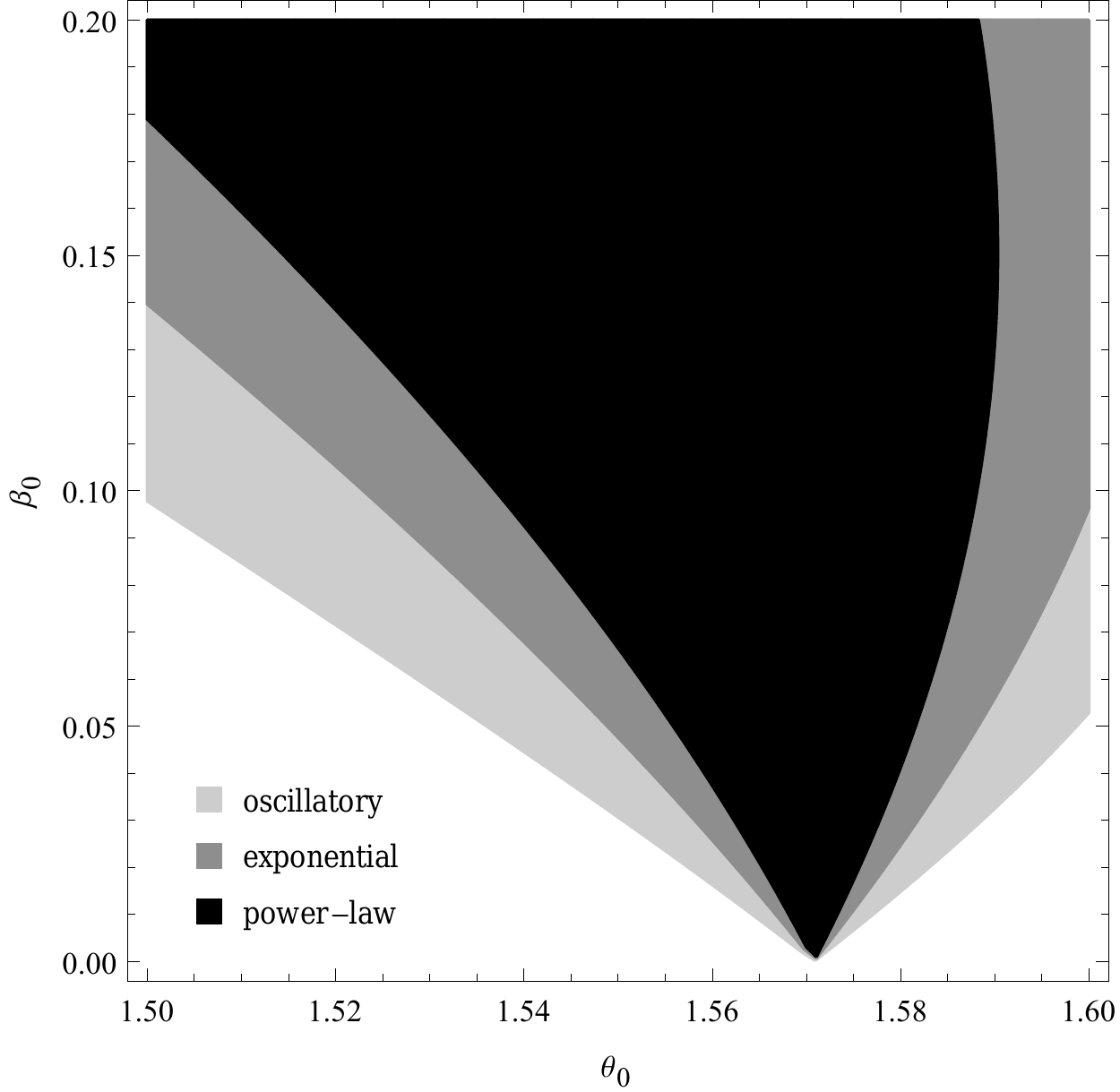}\caption{Plot of the
two-dimensional parametric regions $\mathcal{P}\left(  \beta_{0}\text{,
}\theta_{0}\right)  $ where the oscillatory (light grey), the exponential
decay (grey), and the power law decay (black) driving strategies outperform
the constant driving scheme in terms of both geodesic speed and robustness.
Note that the light grey region contains the grey region while the grey region
contains the black region. Furthermore, the black region specifies a driving
strategy more robust than that specified by the grey region while the grey
region specifies a driving strategy more robust than that specified by the
light grey region. In all region plots, we set $\Gamma/\hslash=1$,
$\lambda=2/\pi$, and $\dot{\theta}_{0}=1$. Finally, the maximum success
probability of each driving scheme is assumed to be greater than
$25/26\simeq\allowbreak0.96$, that is, $0\leq\beta_{0}\lesssim0.20$.}%
\end{figure}

\section{Conclusions}

In this paper, we presented an information geometric analysis of off-resonance
effects on classes of exactly solvable generalized semi-classical Rabi
systems. Specifically, we considered population transfer performed by four
distinct off-resonant driving schemes specified by $\mathrm{su}\left(
2\text{; }%
\mathbb{C}
\right)  $ time-dependent Hamiltonian models. For each scheme, we studied the
consequences of a departure from the on-resonance condition in terms of both
geodesic paths and geodesic speeds on the corresponding manifold of transition
probability vectors. In particular, we analyzed the robustness of each driving
scheme against off-resonance effects. Moreover, we reported on a possible
tradeoff between speed and robustness in the driving schemes being
investigated. Finally, we discussed the emergence of a different relative
ranking in terms of performance among the various driving schemes when
transitioning from the on-resonant to the off-resonant scenarios.

Our main findings can be outlined as follows.

\begin{enumerate}
\item[{[1]}] In the presence of off-resonance effects, the success probability
of the various quantum strategies is affected both in terms of amplitude and
periodicity. In particular, focusing on the constant driving case, the success
probability $p_{0}\left(  \theta\right)  $ in Eq. (\ref{probability1}) is
dampened by the Lorentzian-like factor $1/(1+\beta_{0}^{2})$. Moreover, the
periodicity of the oscillations of this probability changes from $2\pi$ to
$2\pi/\left(  1+\beta_{0}^{2}\right)  ^{\frac{1}{2}}$. In summary,
oscillations become smaller in amplitude but higher in frequency. These facts
are clearly visible in Fig. $1$.

\item[{[2]}] Departing from the on-resonance condition (that is, $\beta_{0}%
=0$), we observe a change in the geodesic paths on the underlying manifolds.
The presence of off-resonance effects (that is, $\beta_{0}\neq0$) generates
geodesic paths with a more complex structure. In particular, unlike what
happens when the on-resonance condition is satisfied, numerical integration of
the geodesic equations is required. In particular, the numerical plots of the
geodesic paths $\theta$ versus the affine parameter $\xi$ for the four driving
strategies with geodesic equations in\ Eqs. (\ref{cumpa1}), (\ref{cumpa2}),
(\ref{cumpa3}), and (\ref{cumpa4}) appear in Fig. $2$.

\item[{[3]}] Each and every numerical value of the geodesic speeds
corresponding to the various quantum driving strategies considered in this
paper becomes smaller when departing from the on-resonance condition (see Fig.
$3$, for instance). In general, quantum driving strategies characterized by a
high geodesic speed appear to be very sensitive to off-resonance effects.
Instead, strategies yielding geodesic paths with smaller geodesic speed values
seem to be more robust against departures from the on-resonance condition.
These observations can be understood from Fig. $4$.

\item[{[4]}] In the off-resonance regime, there emerges a sensitive dependence
of the geodesic speeds on the initial conditions. This, in turn, can cause a
change in the ranking of the various quantum driving strategies. In
particular, the strategy specified by a constant Fisher information is no
longer the absolute best strategy in terms of speed. Indeed, it is possible to
find two-dimensional (2D) parametric regions where this strategy is being
outperformed by all the remaining strategies both in terms of speed and
robustness. In general, these 2D regions are more extended for the less robust
strategies (see Eqs. (\ref{TG1}) and (\ref{TG2})). For instance, we can
identify 2D regions where the speed-based ranking becomes: 1) oscillatory
strategy; 2) exponential-law decay strategy; 3) power-law decay strategy.
Instead, considering the very same 2D regions, the robustness-based ranking is
given by: 1) power-law decay strategy; 2) exponential-law decay strategy; 3)
oscillatory strategy. In these 2D regions, the constant strategy is the worst,
both in terms of speed and robustness. These findings are illustrated in Fig.
$5$.
\end{enumerate}

In conclusion, our information geometric analysis suggests that speed and
robustness appear to be competing features when departing from the
on-resonance condition. Ideally, a quantum driving strategy should be fast,
robust, and thermodynamically efficient \cite{nature,renner18,deffner17}.
Since speed and thermodynamic efficiency quantified in terms of minimum
entropy production paths seem to be conflicting properties as well
\cite{carlopre,carlotechnical19}, it appears reasonable to think there might
be a connection between robustness and thermodynamic efficiency. We leave the
exploration of this conjectured link to future scientific efforts.

\begin{acknowledgments}
C. C. is grateful to the United States Air Force Research Laboratory (AFRL)
Summer Faculty Fellowship Program for providing support for this work. Any
opinions, findings and conclusions or recommendations expressed in this
manuscript are those of the authors and do not necessarily reflect the views
of AFRL.
\end{acknowledgments}

\bigskip\pagebreak

\appendix

\section{Violating a resonance condition}

In this Appendix, we provide two illustrative examples concerning the
violation of a resonance condition.

\subsection{A classical scenario}

In the framework of classical mechanics, one can imagine violating a resonance
condition in a number of manners. For instance, consider a mass-spring system
in the presence of damping and sinusoidal forcing term,%
\begin{equation}
m\ddot{x}+\lambda\dot{x}+kx=f_{0}\cos\left(  \gamma t\right)  \text{.}
\label{HO}%
\end{equation}
In Eq. (\ref{HO}), $\dot{x}\overset{\text{def}}{=}dx/dt$. Furthermore, $m$,
$\lambda$, $k$, and $f_{0}$ denote the mass, the damping coefficient, the
spring constant, and the amplitude of the forcing term, respectively. The
classical resonance curve $\mathcal{C}\left(  \gamma\right)  $,
\begin{equation}
\mathcal{C}\left(  \gamma\right)  \overset{\text{def}}{=}\left[  \left(
k-m\gamma^{2}\right)  ^{2}+b^{2}\gamma^{2}\right]  ^{-1/2}\text{,} \label{rc1}%
\end{equation}
for this specific physical system is proportional to the amplitude
$\mathcal{A}\left(  \gamma\right)  \overset{\text{def}}{=}f_{0}\mathcal{C}%
\left(  \gamma\right)  $ of the steady-state solution to Eq. (\ref{HO}).
Moreover, the resonance condition $\gamma=\gamma^{\ast}$ with%
\begin{equation}
\gamma^{\ast}=\gamma^{\ast}\left(  m\text{, }k\text{, }\lambda\right)
\overset{\text{def}}{=}\left(  \frac{k}{m}-\frac{\lambda^{2}}{2m^{2}}\right)
^{1/2}\text{,} \label{condition1}%
\end{equation}
is specified by imposing the maximum $\mathcal{\bar{C}}\overset{\text{def}}%
{=}\mathcal{C}\left(  \gamma^{\ast}\right)  $ of the resonance curve
$\mathcal{C}\left(  \gamma\right)  $ in Eq. (\ref{rc1}). Therefore, we clearly
note from Eq. (\ref{condition1}) that departures from the resonance condition
$\gamma=\gamma^{\ast}$ can happen by varying the stiffness ($k$), the mass
($m$), and/or the damping ($\lambda$).

\subsection{A quantum scenario}

In the framework of quantum mechanics, one can imagine violating a resonance
condition in a number of manners. For instance, consider a two-level quantum
system described by a time-dependent Hamiltonian,%
\begin{equation}
\mathcal{H}\left(  t\right)  \overset{\text{def}}{=}\mathcal{H}_{0}%
+\mathcal{V}\left(  t\right)  \text{,} \label{hamo}%
\end{equation}
where $\mathcal{H}_{0}$ is the part of the Hamiltonian $\mathcal{H}$ that does
not contain time explicitly while $\mathcal{V}\left(  t\right)  $ is the
time-dependent sinusoidal oscillatory potential. The quantities $\mathcal{H}%
_{0}$ and $\mathcal{V}\left(  t\right)  $ are defined as,%
\begin{equation}
\mathcal{H}_{0}\overset{\text{def}}{=}E_{1}\left\vert 1\right\rangle
\left\langle 1\right\vert +E_{2}\left\vert 2\right\rangle \left\langle
2\right\vert \text{, }\mathcal{V}\left(  t\right)  \overset{\text{def}}%
{=}\gamma e^{i\omega t}\left\vert 1\right\rangle \left\langle 2\right\vert
+\gamma e^{-i\omega t}\left\vert 2\right\rangle \left\langle 1\right\vert
\text{, }%
\end{equation}
respectively, where $\gamma$ and $\omega$ belong to $%
\mathbb{R}
_{+}\backslash\left\{  0\right\}  $. Furthermore, $\left\vert 1\right\rangle $
and $\left\vert 2\right\rangle $ are two eigenstates of $\mathcal{H}_{0}$ with
corresponding eigenvalues $E_{1}$ and $E_{2}$, respectively, with $E_{2}%
>E_{1}$. The quantum resonance curve $\mathcal{Q}\left(  \omega\right)  $,
\begin{equation}
\mathcal{Q}\left(  \omega\right)  \overset{\text{def}}{=}\left[  1+\left(
\frac{\hslash}{\gamma}\right)  ^{2}\frac{\left(  \omega-\omega_{21}\right)
^{2}}{4}\right]  ^{-1}\text{,} \label{rc2}%
\end{equation}
for this specific physical system described by the Hamiltonian in Eq.
(\ref{hamo}) is the amplitude squared that characterizes the transition
probability $\mathcal{P}_{\left\vert 1\right\rangle \rightarrow\left\vert
2\right\rangle }$ (assuming that at $t=0$ only the level $\left\vert
1\right\rangle $ is populated) between the two quantum states $\left\vert
1\right\rangle $ and $\left\vert 2\right\rangle $ with $\omega_{21}%
\overset{\text{def}}{=}\left(  E_{2}-E_{1}\right)  /\hslash$ being the
characteristic frequency of the system. Moreover, the resonance condition
$\omega=\omega^{\ast}$ with%
\begin{equation}
\omega^{\ast}=\omega_{21}=\omega_{21}\left(  E_{1}\text{, }E_{2}\right)
\overset{\text{def}}{=}\left(  E_{2}-E_{1}\right)  /\hslash\text{,}
\label{condition2}%
\end{equation}
is specified by imposing the maximum $\mathcal{\bar{Q}}\overset{\text{def}}%
{=}\mathcal{Q}\left(  \omega_{21}\right)  $ of the resonance curve
$\mathcal{Q}\left(  \omega\right)  $ in Eq. (\ref{rc2}). Therefore, we clearly
note from Eq. (\ref{condition2}) that departures from the resonance condition
$\omega=\omega^{\ast}$ can happen by varying the larger ($E_{2}$) and the
smaller ($E_{1}$) energy levels. We point out that considering the following
correspondences%
\begin{equation}
\omega\rightarrow\omega\text{, }\gamma\rightarrow\frac{\left\vert e\right\vert
\hslash B_{\perp}}{2mc}\text{, and }\omega_{21}\rightarrow\frac{\left\vert
e\right\vert B_{\parallel}}{mc}\text{,}%
\end{equation}
it can be shown that the Hamiltonian in Eq. (\ref{hamo}) describes a two-level
quantum mechanical system represented by a spin-$1/2$ particle immersed in an
external magnetic field $\vec{B}\left(  t\right)  $ given by,%
\begin{equation}
\vec{B}\left(  t\right)  \overset{\text{def}}{=}B_{\parallel}\hat{z}+B_{\perp
}\left[  \cos\left(  \omega t\right)  \hat{x}+\sin\left(  \omega t\right)
\hat{y}\right]  \text{.}%
\end{equation}
The resonance condition $\omega=\omega^{\ast}$ is satisfied whenever the
positively assumed frequency $\omega$ of the rotating magnetic field in the
$xy$-plane equals Larmor's precessional frequency specified by the intensity
of the uniform magnetic field $B_{\parallel}\hat{z}$ \cite{sakurai},%
\begin{equation}
\omega^{\ast}=\omega^{\ast}\left(  \left\vert e\right\vert \text{,
}B_{\parallel}\text{, }m\text{, }c\text{ }\right)  \overset{\text{def}}%
{=}\frac{\left\vert e\right\vert B_{\parallel}}{mc}\text{.}%
\end{equation}
Finally, we emphasize that the analogue of the adimensional parameter
$\beta_{0}$ used throughout our paper becomes in this quantum mechanical
framework the quantity defined as%
\begin{equation}
\beta_{0}\overset{\text{def}}{=}\frac{mc}{\left\vert e\right\vert B_{\perp}%
}\left(  \omega-\omega^{\ast}\right)  =\frac{mc}{\left\vert e\right\vert
B_{\perp}}\left(  \omega-\frac{\left\vert e\right\vert B_{\parallel}}%
{mc}\right)  \text{.} \label{staticbetazero}%
\end{equation}
Observe that Eq. (\ref{staticbetazero}) is the static version of Eq.
(\ref{betazero}). For further details on the quantum mechanical motion in
two-level quantum systems described by time-dependent Hamiltonians, we refer
to Ref. \cite{sakurai}.

\section{Speed of geodesics}

In the Appendix, we briefly show that geodesics have constant speed. For
further details, we refer to Ref. \cite{lee97}.

Recall that if $\Theta\left(  \xi\right)  \overset{\text{def}}{=}\left(
\theta^{1}\left(  \xi\right)  \text{,..., }\theta^{n}\left(  \xi\right)
\right)  $ is a curve in a $n$-dimensional Riemannian manifold $\mathcal{M}$,
the speed of $\Theta$ at any time is the length of its velocity vector
$\left\Vert \dot{\Theta}\left(  \xi\right)  \right\Vert $,%
\begin{equation}
\left\Vert \dot{\Theta}\left(  \xi\right)  \right\Vert \overset{\text{def}}%
{=}\left\langle \dot{\Theta}\left(  \xi\right)  \text{, }\dot{\Theta}\left(
\xi\right)  \right\rangle ^{\frac{1}{2}}=\left[  g_{ab}\left(  \theta\right)
\dot{\theta}^{a}\dot{\theta}^{b}\right]  ^{\frac{1}{2}}\text{,} \label{speed}%
\end{equation}
where $g_{ab}\left(  \theta\right)  $ denotes a Riemannian metric on the
manifold $\mathcal{M}$ while $\left\{  \theta^{a}\left(  \xi\right)  \right\}
$ are the component functions of the curve $\Theta\left(  \xi\right)  $. We
say $\Theta\left(  \xi\right)  $ is constant speed if $\left\Vert \dot{\Theta
}\left(  \xi\right)  \right\Vert $ in\ Eq. (\ref{speed}) does not depend on
$\xi$, and unit speed if the speed is identically equal to one.

It is well-known that all Riemannian geodesics are constant speed curves.
Indeed, let $\mathcal{M}$ be a manifold with a linear connection $\nabla$, and
let $\Theta$ be a curve on $\mathcal{M}$. The acceleration of $\Theta$ is the
vector field $D_{\xi}\dot{\Theta}$ along $\Theta$ (that is, the covariant
derivative of $\dot{\Theta}\left(  \xi\right)  $ along $\Theta$). A curve
$\Theta$ is called a geodesic with respect to $\nabla$ if its acceleration
$D_{\xi}\dot{\Theta}$ equals zero (that is, the velocity vector field
$\dot{\Theta}$ is parallel along the curve $\Theta$). Therefore, for any
geodesic $\Theta\left(  \xi\right)  $, we have%
\begin{align}
\frac{d}{d\xi}\left[  \left\Vert \dot{\Theta}\left(  \xi\right)  \right\Vert
^{2}\right]   &  =\frac{d}{d\xi}\left[  \left\langle \dot{\Theta}\left(
\xi\right)  \text{, }\dot{\Theta}\left(  \xi\right)  \right\rangle \right]
\nonumber\\
&  =\left\langle D_{\xi}\dot{\Theta}\left(  \xi\right)  \text{, }\dot{\Theta
}\left(  \xi\right)  \right\rangle +\left\langle \dot{\Theta}\left(
\xi\right)  \text{, }D_{\xi}\dot{\Theta}\left(  \xi\right)  \right\rangle
\nonumber\\
&  =0\text{,}%
\end{align}
that is,%
\begin{equation}
\frac{d}{d\xi}\left[  \left\Vert \dot{\Theta}\left(  \xi\right)  \right\Vert
^{2}\right]  =0\text{.} \label{finalino}%
\end{equation}
From Eq. (\ref{finalino}), we conclude that geodesics $\left\{  \Theta\left(
\xi\right)  \right\}  $ have constant speed $\left\{  \left\Vert \dot{\Theta
}\left(  \xi\right)  \right\Vert \right\}  $.

\end{document}